\documentclass[lettersize,journal]{IEEEtran}
\usepackage{amsmath,amsfonts}
\usepackage{array}
\usepackage[caption=false,font=normalsize,labelfont=sf,textfont=sf]{subfig}
\usepackage{textcomp}
\usepackage{stfloats}
\usepackage{url}
\usepackage{verbatim}
\usepackage{graphicx}
\usepackage{cite}
\usepackage{nicematrix}
\usepackage[toc,acronym]{glossaries}
\usepackage{hyperref}
\usepackage{booktabs}
\usepackage{algorithmic}
\usepackage[ruled,vlined]{algorithm2e}
\usepackage{tikz}
\usetikzlibrary{calc}
\usetikzlibrary{patterns, decorations.markings}
\usepackage{xcolor}
\usepackage{pifont}%
\newcommand{\cmark}{\ding{51}}%
\newcommand{\xmark}{\ding{55}}%
\newcommand{\highlight}[1]{\textcolor{black}{#1}}
\hypersetup{
    colorlinks=true,
    linkcolor=blue,
    filecolor=magenta,
    urlcolor=cyan,
    }

\def\BibTeX{{\rm B\kern-.05em{\sc i\kern-.025em b}\kern-.08em
    T\kern-.1667em\lower.7ex\hbox{E}\kern-.125emX}}
\usepackage{balance}

\newacronym{6lowpan}{6LoWPAN}{IPv6 over Low-Power Wireless Personal Area Networks}
\newacronym{ai}{AI}{Artificial Intelligence}
\newacronym{ap}{AP}{Access Point}
\newacronym{api}{API}{Application Program Interface}
\newacronym{ae}{AE}{Autoencoder}
\newacronym{automl}{AutoML}{Automated Machine Learning}
\newacronym{asn}{ASN}{Absolute Slot Number}
\newacronym{bfs}{BFS}{Breadth First Search}
\newacronym{blip}{BLIP}{Berkeley Low-power IP}
\newacronym{ctp}{CTP}{Collection Tree Protocol}
\newacronym{cnn}{CNN}{Convolutional Neural Networks}
\newacronym{dt}{DT}{Decision Tree}
\newacronym{dl}{DL}{Deep Learning}
\newacronym{dqn}{DQN}{Deep Q-Learning}
\newacronym{drl}{DRL}{Deep Reinforcement Learning}
\newacronym{eos}{EOS}{Embedded Operating System}
\newacronym{eb}{EB}{Enhanced Beacon}
\newacronym{ewma}{EWMA}{Exponential Weighted Moving Average}
\newacronym{fpga}{FPGA}{Field-Programmable Gate Array}
\newacronym{fsm}{FSM}{Finite State Machine}
\newacronym{gan}{GAN}{Generative Adversarial Network}
\newacronym{gprs}{GPRS}{General Packet Radio Service}
\newacronym{hrl}{HRL}{Hierarchical Reinforcement Learning}
\newacronym{hrltsch}{HRL-TSCH}{Hierarchical Reinforcement Learning-based Time Slotted Channel Hopping}
\newacronym{ia}{IA}{Intelligent Agent}
\newacronym{icmpv6}{ICMPv6}{Internet Control Message Protocol version 6}
\newacronym{ietf}{IETF}{Internet Engineering Task Force}
\newacronym{iot}{IoT}{Internet of Things}
\newacronym{iiot}{IIoT}{Industrial Internet of Things}
\newacronym{ids}{IDS}{Intrusion Detection System}
\newacronym{ip}{IP}{Internet Protocol}
\newacronym{ipv4}{IPv4}{Internet Protocol version 4}
\newacronym{ipv6}{IPv6}{Internet Protocol version 6}
\newacronym{knn}{k-NN}{K-Nearest Neighbour}
\newacronym{kpi}{KPI}{Key Performance Indicator}
\newacronym{lan}{LAN}{Local Area Network}
\newacronym{wlan}{WLAN}{Wireless Local Area Network}
\newacronym{lora}{LoRa}{Long Range}
\newacronym{lorawan}{LoRaWAN}{Long Range Wide Area Network}
\newacronym{lowpan}{LoWPAN}{Low-Power Wireless Personal Area Networks}
\newacronym{lqi}{LQI}{Link Quality Indicator}
\newacronym{m2m}{M2M}{Machine-to-Machine}
\newacronym{mac}{MAC}{Media Access Control}
\newacronym{mdp}{MDP}{Markov Decision Process}
\newacronym{mems}{MEMS}{Micro-Electro-Mechanical Systems}
\newacronym{mcu}{MCU}{Microcontroller Unit}
\newacronym{ml}{ML}{Machine Learning}
\newacronym{mlsdwsn}{ML-SDWSN}{Machine Learning Software-Defined Wireless Sensor Network}
\newacronym{msf}{MSF}{Minimal Scheduling Function}
\newacronym{mst}{MST}{Minimum Spanning Tree}
\newacronym{na}{NA}{Neighbor Advertisement}
\newacronym{nc}{NC}{Network Configuration}
\newacronym{nes}{NES}{Networked Embedded Systems}
\newacronym{nd}{ND}{Neighbor Discovery}
\newacronym{nl}{NL}{Network Lifetime}
\newacronym{nlp}{NLP}{Non-Linear Programming}
\newacronym{nn}{NN}{Neural Network}
\newacronym{os}{OS}{Operating System}
\newacronym{pdr}{PDR}{Packet Delivery Ratio}
\newacronym{plr}{PLR}{Packet Loss Rate}
\newacronym{pso}{PSO}{Particle Swarm Optimisation}
\newacronym{pca}{PCA}{Principal Component Analysis}
\newacronym{qos}{QoS}{Quality of Service}
\newacronym{ql}{QL}{Q-Learning}
\newacronym{ram}{RAM}{Random-Access Memory}
\newacronym{rdc}{RDC}{Radio Duty-Cycle}
\newacronym{rom}{ROM}{Read-Only Memory}
\newacronym{rnn}{RNN}{Recurrent Neural Network}
\newacronym{rl}{RL}{Reinforcement Learning}
\newacronym{rtt}{RTT}{Round-Trip Time}
\newacronym{rpl}{RPL}{Routing Protocol for Low-Power and Lossy Networks}
\newacronym{rssi}{RSSI}{Received Signal Strength Indicator}
\newacronym{stp}{STP}{Spanning Tree Protocol}
\newacronym{svm}{SVM}{Support Vector Machine}
\newacronym{sdn}{SDN}{Software-Defined Networking}
\newacronym{snr}{SNR}{Signal to Noise Ratio}
\newacronym{slip}{SLIP}{Serial Line Internet Protocol}
\newacronym{sdwsn}{SDWSN}{Software-Defined Wireless Sensor Network}
\newacronym{sp}{SP}{Shortest Path}
\newacronym{tl}{TL}{Transfer Learning}
\newacronym{tlm}{TLM}{Traffic Load Minimisation}
\newacronym{tcp}{TCP}{Transmission Control Protocol}
\newacronym{tsch}{TSCH}{Time Slotted Channel Hopping}
\newacronym{udp}{UDP}{User Datagram Protocol}
\newacronym{udgm}{UDGM}{Unit Disk Graph Medium}
\newacronym{uip}{$\mu$IP}{micro Internet Protocol}
\newacronym{uipv6}{$\mu$IPv6}{micro Internet Protocol version 6}
\newacronym{wam}{WAM}{Weighted Arithmetic Mean}
\newacronym{wban}{WBAN}{Wireless Body Area Network}
\newacronym{wpan}{WPAN}{Wireless Personal Area Network}
\newacronym{wsan}{WSAN}{Wireless Sensor and Actuator Network}
\newacronym{wsn}{WSN}{Wireless Sensor Network}
\newacronym{sn}{SN}{Sensor Node}

\begin{document}
\title{HRL-TSCH: A Hierarchical Reinforcement Learning-based TSCH Scheduler for IIoT}
\author{F. Fernando Jurado-Lasso, \IEEEmembership{Member, IEEE}, Charalampos Orfanidis, \IEEEmembership{Member, IEEE}, J. F. Jurado, and Xenofon Fafoutis, \IEEEmembership{Senior Member, IEEE}
    \thanks{Manuscript received February 1, 2024; revised xx, xx. This work was partly
        supported by DAIS. DAIS (https://dais-project.eu/) has received funding from
        the ECSEL Joint Undertaking (JU) under grant agreement No 101007273.
        The JU receives support from the European Union's Horizon 2020 research
        and innovation programme and Sweden, Spain, Portugal, Belgium, Germany,
        Slovenia, Czech Republic, Netherlands, Denmark, Norway, and Turkey. The
        document reflects only the authors' view, and the Commission is not responsible
        for any use that may be made of the information it contains. Danish
        participants are supported by Innovation Fund Denmark under grant agreement
        No. 0228-00004A. \textit{(Corresponding author: F. Fernando Jurado-Lasso)}}
    \thanks{F. Fernando Jurado-Lasso, Charalampos Orfanidis, and Xenofon Fafoutis are with the
        Embedded Systems Engineering section, DTU Compute,
        Technical University of Denmark, 2800 Lyngby, Denmark (e-mail:
        ffjla@dtu.dk; chaorf@dtu.dk; xefa@dtu.dk).}
    \thanks{J. F. Jurado is with the Department of Basic Science, Faculty of Engineering
        and Administration, Universidad Nacional de Colombia Sede Palmira, Palmira
        763531, Colombia (e-mail: jfjurado@unal.edu.co).}}

\markboth{Journal of \LaTeX\ Class Files,~Vol.~18, No.~9, February~2024}%
{How to Use the IEEEtran \LaTeX \ Templates}

\maketitle

\begin{abstract}
    The \acrfull{iiot} demands adaptable \acrfull{nes} for optimal performance.
Combined with recent advances in \acrfull{ai}, tailored solutions can be developed to meet specific application requirements.
This study introduces \acrshort{hrltsch}, an approach rooted in \acrfull{hrl}, to devise \acrfull{tsch} schedules provisioning \acrshort{iiot} demand.
\acrshort{hrltsch} employs dual policies: one at a higher level for \acrshort{tsch} schedule link management, and another at a lower level for timeslot and channel assignments.
The proposed \acrshort{rl} agents address a multi-objective problem, optimizing throughput, power efficiency, and network delay based on predefined application requirements.
Simulation experiments demonstrate \acrshort{hrltsch}'s superiority over existing state-of-art approaches, effectively achieving an optimal balance between throughput, power consumption, and delay, thereby enhancing \acrshort{iiot} network performance.

\end{abstract}


\section{Introduction}
\label{sec:introduction}
\IEEEPARstart{T}{he} \acrfull{iot} is a transformative technology that interconnects objects to the Internet, paving the way for innovative applications and services~\cite{he2022collaborative}.
These objects equipped with sensors, processing capabilities, power sources, and wireless communication radios, form the backbone of the \acrfull{iiot}.
The \acrshort{iiot} specifically leverages \acrshort{iot} devices to facilitate real-time control and monitoring of industrial processes, enhancing operational efficiency and decision-making \cite{da2014internet}.

In the realm of \acrshort{iiot} applications, diverse needs arise, ranging from data rate and delay to throughput and power usage \cite{sisinni2018industrial}.
Some applications demand energy-efficient modes to extend the lifespan of sensor nodes, particularly in remote and challenging environments.
Others prioritize low latency for real-time industrial control, while some strike a fair balance between reliability and power efficiency.
As operational needs evolve, the network must dynamically adapt to ensure seamless performance.

\acrfull{nes} traditionally serve as the foundation for both \acrshort{iot} and \acrshort{iiot}.
These networks consist of numerous low-cost, low-power wireless sensor nodes (often called \acrfullpl{sn}) deployed in the environment to monitor and control physical phenomena \cite{yick2008wireless}.
In challenging environments susceptible to interference and multipath fading, a conventional approach involves utilizing \acrfull{tsch}, a \acrfull{mac} protocol of the IEEE 802.15.4 standard.
\acrshort{tsch} schedules packet transmission and reception in a time-slotted fashion, overcoming the challenges of interference and multipath fading \cite{duquennoy2017tsch}.

The effectiveness of a \acrshort{tsch} network heavily relies on its scheduler, responsible for assigning timeslots and channels to communication links.
While redundant links enhance network reliability and minimize packet delay, they may also increase power consumption.
Therefore, the \acrshort{tsch} scheduler must be cautiously designed to meet the requirements of \acrshort{iiot} applications, ensuring flexibility and adaptability to dynamic changes.

In this context, this paper introduces Hierarchical Reinforcement Learning for Time-Slotted Channel Hopping (\acrshort{hrltsch}).
This innovative approach, grounded in \acrfull{hrl}, is intricately designed to customize \acrshort{tsch} schedules according to the specific requirements of \acrshort{iiot} applications.
The hierarchical framework comprises \acrshort{rl} agents responsible for learning the optimal schedule and a \acrshort{tsch} selection algorithm enabling \acrshortpl{sn} to efficiently select the nearest scheduled link for a designated destination address.
The incorporation of \acrshort{hrl} proves highly beneficial as it efficiently manages the inherent complexity of the \acrshort{tsch} link scheduling problem.
By partitioning decision-making into higher and lower levels, our approach optimizes network performance by considering both global changes and local link-specific policies.
This hierarchical structure significantly enhances adaptability, efficiency in exploration and exploitation, scalability, and generalization capabilities.
The proposed methodology contributes to the field by providing a flexible and responsive solution that maximizes network performance in dynamic \acrshort{iiot} environments.
In brief terms, the contributions of this paper are as follows:

\begin{enumerate}
    \item We develop a \acrshort{hrl} architecture specifically designed for solving the \acrshort{tsch} scheduler problem in \acrshort{iiot} networks that aims to maximize network performance.
    \item We formulate of a comprehensive mathematical model for estimating power consumption, delay, and throughput within a \acrshort{tsch} network.
          This model serves as the foundation for the \acrshort{rl} agent's learning process to derive the optimal schedule.
    \item We develop a \acrshort{tsch} selection algorithm for end devices to efficiently select the nearest scheduled link corresponding to a specific destination address.
    \item We conduct a comprehensive performance evaluation of the proposed approach using various \acrshort{iiot} application requirements within the Cooja network simulator.
\end{enumerate}

The rest of the paper is organized as follows. Section \ref{sec:background} provides an introduction to key concepts and technologies relevant to this study.
\highlight{In Section \ref{sec:related_work}, we review related work, focusing on centralized and decentralized schedulers. We highlight the limitations of existing centralized approaches in adapting to dynamic changes, underscoring the need for flexibility. This informs our analysis and the development of our proposed framework.}
Section \ref{sec:methodology} offers a comprehensive overview of the methodology employed in designing the \acrshort{tsch} scheduler.
Section \ref{sec:reinforcement_learning} focuses on the \acrshort{rl} approach utilized in the design of the \acrshort{tsch} scheduler.
Section \ref{sec:performance_evaluation} presents the performance evaluation of the proposed approach, showcasing its effectiveness and efficiency.
Lastly, Section \ref{sec:conclusion} concludes the paper and provides insights into future directions and potential avenues for further research.

\section{Background}\label{sec:background}

\subsection{Time Slotted Channel Hopping (TSCH)}
The \acrshort{tsch} protocol operates at the link-layer and is characterized by two key features: time synchronization of nodes and frequency hopping functions~\cite{duquennoy2017tsch}.
These components work in tandem to facilitate \textit{frequency-division multiple access} and \textit{time-division multiple access}, enabling multiple network nodes to efficiently share the same radio medium by dividing its available bandwidth into frequency sub-channels.

Allocation of access in discrete timeslots within each sub-channel frequency is managed to ensure equitable sharing among nodes.
\highlight{In centralized schedulers}, a designated \textit{coordinator} node assumes the responsibility of generating a \textit{schedule} based on these rules.
This schedule is then disseminated to all nodes in the network.

To join the \acrshort{tsch} network, every \acrshort{sn} must receive and adhere to this schedule, which defines the \textit{slotframe} (i.e., the set of timeslots) and the \textit{channel hopping sequence} (i.e., the set of frequency sub-channels) to be used by the network.
In essence, the \acrshort{tsch} schedule comprises a set of links, specifying the actions each node must take at designated timeslots.

\acrshortpl{sn} within the network have the option of receiving or transmitting packets or entering a sleep state to conserve energy.
The organization of nodes' transmissions and receptions in the schedule, along with the slotframe size, are critical parameters that significantly influence the performance of the \acrshort{tsch} network.
These factors impact throughput, delay, and energy efficiency.

\subsection{Reinforcement Learning (RL)}
\acrshort{rl} is a branch of \acrshort{ml} that excels at solving complex problems across various domains like robotics and games~\cite{chen2016reinforcement}.
In \acrshort{rl}, an agent learns the best policy $\pi$ by interacting with the environment, aiming to maximize the cumulative reward $\mathcal{R}$~\cite{luong2019applications}.
At each time step $t$, the agent takes action $a_t \in \mathcal{A}$, and the environment responds with a reward $r_{t+1} \in \mathcal{R}$ and a new state $s_{t+1} \in \mathcal{S}$.
The agent's goal is to learn the optimal policy $\pi$ that maximizes the cumulative reward $\mathcal{R}$~\cite{yu2022learning}.
There are two main approaches to \acrshort{rl}: value-based and policy-based.
In value-based \acrshort{rl}, the agent learns the optimal policy by estimating the value function $V(s)$ or $Q(s,a)$.
In policy-based \acrshort{rl}, the agent learns the optimal policy directly without estimating the value function.
In this paper, we use the \acrfull{dqn} algorithm, a value-based \acrshort{rl} algorithm, to train the \acrshort{rl} agents.

\subsubsection{Deep Q-Network (DQN)}
Since our \acrshort{hrl} framework is a model-free approach, we use the \acrshort{dqn} algorithm to train the \acrshort{rl} agents.
The \acrshort{dqn} algorithm is a variant of the Q-learning algorithm that uses a deep neural network to approximate the action-value function $Q(s, a)$, and a replay buffer to store the experiences and then sample from it to train the agent.
Q-learning updates the state-action value function $Q(s, a)$ through an iterative process.
It employs a Bellman equation as the basis for its updates.
The value of $Q(s, a) \gets (1-\sigma) Q(s, a) + \sigma \left[ r + \lambda \max_{a' \in \mathcal{A}} Q(s', a') \right]$ is updated by taking a weighted sum of the previous Q-value and new information obtained from the current state and action.
Where $\sigma \in [0, 1]$ and $\lambda \in [0, 1]$ are the learning rate and discount factor, respectively.
$s'$ and $a'$ are the next state and action, respectively.
$r$ is the immediate reward of the policy $\pi$.

\acrshort{dqn} approximates the Q-value function $Q(s, a)$ using a deep neural network such that $Q(s, a) \approx Q(s, a; \theta)$, where $\theta$ is the set of parameters of the neural network.
The neural network is trained to minimize the loss function $L(\theta) = \mathop{\mathbb{E}}_{\tau \sim U(.)} \left[ \left( \phi - Q(s, a; \theta) \right)^{2} \right]$, where $\phi = r + \lambda \max_{a' \in \mathcal{A}} Q(s', a'; \theta^{-})$
Where $\tau$ is the experience tuple $(s, a, r, s')$, and $U(.)$ is the uniform distribution over all possible experience tuples.
$\theta^{-}$ is the set of parameters of the target network.
The target network is a copy of the neural network that is used to calculate the target Q-value.

\subsection{Hierarchical Reinforcement Learning (HRL)}
\acrshort{hrl} is a branch of \acrshort{rl} that aims to solve complex problems by decomposing them into smaller sub-problems~\cite{barto2003recent}.
In \acrshort{hrl}, there are two levels of policies: a higher-level policy $\pi_h$ and a lower-level policy $\pi_l$.
The higher-level policy $\pi_h$ selects the sub-goal $g$ that the agent should achieve.
The lower-level policy $\pi_l$ selects the action $a$ that the agent should take to achieve the sub-goal $g$.
The higher-level policy $\pi_h$ is trained using the lower-level policy $\pi_l$.
The lower-level policy $\pi_l$ is trained using the reward function of the higher-level policy $\pi_h$.
In this paper, we use the \acrshort{hrl} framework to design the \acrshort{tsch} scheduler.
The higher-level policy $\pi_h$ selects the link that the agent should schedule.
The lower-level policy $\pi_l$ selects the timeslot and channel that the agent should assign to the selected link.
For further insights into \acrshort{hrl}, readers can refer to~\cite{pateria2021hierarchical}.
Additionally, those interested in applications of \acrshort{ml} in \acrshort{nes} can explore~\cite{jurado2022survey, kim2020machine}.

\subsection{\highlight{\acrshort{tsch} and \acrfull{rl}}}

\highlight{In the realm of \acrshort{rl}, the \acrshort{tsch} schedule can be linked to a \textit{policy} dictating the actions of network nodes.
This policy serves as a mapping function, correlating the current state of the network with the appropriate action to undertake.
The network state encompasses variables such as the current timeslot, channel, and the status of network links. Within the framework of \acrshort{rl}, the coordinator assumes the role of an \textit{agent}, tasked with learning the optimal policy through interactions with the environment.}

\section{Related Work}\label{sec:related_work}

\begin{table*}[ht!]
    \centering
    \caption{Summary and comparison of related work.}
    \label{tab:comparison}
    \def\arraystretch{1}%
    \begin{NiceTabular}[c]{cccccm{2cm}m{6.4cm}}[hvlines]
        \CodeBefore
        \rowcolor{lightgray}{1}
        \rowcolors{2}{gray!12}{}[respect-blocks]
        \Body
        \RowStyle[]{\bfseries}
        Scheduler                         & Architecture              & \acrshort{ml}  & Req.               & Dyn. schedule      & Optimization                         & key features                                                                                                               \\
        \cite{jin2016centralized}         & Centralized               & \xmark         & \xmark             & \xmark             & \acrshort{tsch} schedule             & Sequential slot reservation algorithm that minimizes the packet delay in multihop \acrshort{tsch} networks.                \\
        \cite{jerbi2022msu}               & Centralized               & \xmark         & \xmark             & \cmark             & \acrshort{tsch} schedule             & Dynamic scheduling algorithm for mobility in \acrshort{tsch} networks.                                                     \\
        \cite{tavakoli2018topology}       & Centralized               & \xmark         & \xmark             & \xmark             & \acrshort{tsch} schedule             & High timeslot utilization algorithm that minimizes the latency.                                                            \\
        \cite{ojo2018throughput}          & Centralized               & \xmark         & \xmark             & \xmark             & \acrshort{tsch} schedule             & Throughput and max-min fairness scheduling algorithm.                                                                      \\
        \cite{jurado2023elise}            & Centralized               & \acrshort{rl}  & \cmark             & \cmark             & \acrshort{tsch} Slotframe size       & \acrshort{rl} approach that optimizes the slotframe size of \acrshort{tsch} networks.                                      \\
        \cite{nguyen2019rl}               & Centralized               & \acrshort{rl}  & \xmark             & \xmark             & \acrshort{tsch} schedule             & \acrshort{rl} approach that optimizes the Minimal Scheduling Mechanism.                                                    \\
        \cite{dakdouk2018reinforcement}   & Centralized               & \acrshort{rl}  & \xmark             & \xmark             & \acrshort{tsch} schedule             & \acrshort{rl} approach that optimizes the channel selection.                                                               \\
        \cite{veisi2023enabling}          & Centralized               & \xmark         & \xmark             & \cmark             & \acrshort{tsch} schedule             & A Proof-of-concept \acrshort{sdn}-based scheduling algorithm to enable centralized scheduling in \acrshort{tsch} networks. \\
        \cite{kotsiou2019whitelisting}    & Centralized               & \xmark         & \xmark             & \xmark             & \acrshort{tsch} schedule             & A scheduling algorithm that uses a whitelist to avoid collisions.                                                          \\
        \cite{hajizadeh2023decentralized} & Decentralized             & \acrshort{drl} & \xmark             & \xmark             & \acrshort{tsch} parameters           & \acrshort{drl} framework that configures the parameters of \acrshort{tsch} networks.                                       \\
        \highlight{\cite{park2020multi}}  & \highlight{Decentralized} & \highlight{RL} & \highlight{\xmark} & \highlight{\cmark} & \highlight{\acrshort{tsch} schedule} & \highlight{Multi-agent Q-learning approach that optimizes slot allocation in \acrshort{tsch} networks.}                    \\
        \cite{duquennoy2015orchestra}     & Decentralized             & \xmark         & \xmark             & \xmark             & \acrshort{tsch} schedule             & Autonomous scheduling algorithm that builds its own schedule.                                                              \\
        \cite{ha2022traffic}              & Decentralized             & \xmark         & \xmark             & \xmark             & \acrshort{tsch} schedule             & Traffic-aware scheduling algorithm that facilitates load balancing.                                                        \\
        \cite{farag2020rea}               & Decentralized             & \xmark         & \xmark             & \xmark             & \acrshort{tsch} schedule             & Emergency-aware scheduling algorithm that forwards emergency traffic with high reliability and bounded delay.              \\
        \cite{bommisetty2022resource}     & Decentralized             & \xmark         & \xmark             & \xmark             & \acrshort{tsch} schedule             & \acrshort{drl}-based scheduling algorithm that implements the scheduling policy as an optimization problem.                \\
        \RowStyle[]{\bfseries}
        \acrshort{hrltsch}                & Centralized               & \acrshort{hrl} & \cmark             & \cmark             & \acrshort{tsch} schedule             & \acrshort{hrl} framework that optimizes the \acrshort{tsch} schedule based on the application requirements.                \\
    \end{NiceTabular}
\end{table*}

\acrshort{tsch} schedulers fall into four categories: centralized, decentralized, static, and hybrid~\cite{urke2021survey,jerbi2022msu}.
In this section, we focus on centralized and decentralized schedulers, given their relevance to our work.
\subsection{Centralized Schedulers}

Centralized schedulers are designed by a scheduling algorithm that runs on a central entity, such as a controller or a border router, that has a global view of the network.
For instance, in~\cite{jin2016centralized} the authors propose a scheduling algorithm that reserves sequential slots along the path from the source to the destination, minimizing packet delay in multihop \acrshort{tsch} networks.
However, the algorithm complexity increases with the network size.
Dynamic scheduling is addressed in \cite{jerbi2022msu}, which proposes an algorithm considering end-to-end delay, network throughput, and \acrfull{pdr} to handle mobility in \acrshort{tsch} networks.
Despite its effectiveness, tracking node mobility introduces significant overhead, and dynamic adaptation to the traffic pattern is not considered.
A cross-layer approach optimizing topology and \acrshort{tsch} schedule is presented in \cite{tavakoli2018topology} to minimize latency.
While achieving high timeslot utilization, power consumption remains a concern.
\cite{ojo2018throughput} proposes a scheduling algorithm aiming to maximize network throughput and ensure max-min fairness.
However, prioritizing throughput may lead to increased power consumption.
\cite{jurado2023elise} adopts a \acrshort{rl}-based approach to optimize slotframe size in \acrshort{tsch} networks.
While slotframe size is optimized based on specific application requirements, the \acrshort{tsch} schedule itself is not optimized.
Optimizing the Minimal Scheduling Mechanism using \acrshort{rl} and \acrfull{mdp} is proposed by Nguyen-Duy et al.~\cite{nguyen2019rl}, where nodes keep their radios activated based on traffic patterns from various application scenarios.
However, dynamic adaptation of the schedule is not considered.
\cite{dakdouk2018reinforcement} evaluates nine Multi-Armed Bandit (MAB) algorithms to select the optimal channel and introduces a mechanism integrating selected algorithms with \acrshort{tsch} to improve reliability and energy efficiency.
Nevertheless, the mechanism lacks dynamic adaptation to traffic patterns.
Other noteworthy centralized schedulers include \cite{veisi2023enabling}, introducing a scheduling algorithm rooted in \acrfull{sdn} for centralized scheduling in \acrshort{tsch} networks, and \cite{kotsiou2019whitelisting}, proposing a scheduling algorithm using a whitelist to avoid collisions.

\subsection{Decentralized Schedulers}

The decentralized schedulers include autonomous approaches that allow \acrshortpl{sn} to autonomously select the timeslots and channels to transmit and receive packets and collaborative approaches that require \acrshortpl{sn} to collaborate to design the schedule.
An example is presented in \cite{hajizadeh2023decentralized}, where a \acrfull{drl} framework is employed to offer \acrfull{qos} features.
However, this framework lacks awareness of changes in application requirements and primarily focuses on the parameter configuration of the \acrshort{tsch} protocol.
\highlight{
    In~\cite{park2020multi}, QL-TSCH is proposed, an autonomous scheduling algorithm that uses a multi-agent Q-learning approach to optimize slot allocation in \acrshort{tsch} networks.
    Each agent is responsible for a node and learns the optimal transmission slot based on the network conditions.
    The algorithm uses an action peeking mechanism, which allows nodes to keep track of the actions of their neighbors by constantly monitoring the time slots, to reduce learning time and improve convergence rate.
    However, the algorithm suffers from high energy consumption due to constant monitoring of the time slots.
}
In \cite{duquennoy2015orchestra}, an autonomous scheduling algorithm is introduced, allowing each node to construct its schedule without negotiation overhead.
Nevertheless, the Orchestra schedule may lead to suboptimal network performance due to its lack of consideration for network conditions.
A traffic-aware scheduling algorithm for 6TiSCH networks is proposed in \cite{ha2022traffic}.
This algorithm utilizes cell allocation information to enhance load balancing and improve bandwidth utilization.
However, it lacks the flexibility to enable run-time reconfiguration of the schedule.
Addressing emergency scenarios, \cite{farag2020rea} presents an emergency-aware scheduling algorithm for \acrshort{tsch} networks.
This algorithm prioritizes emergency traffic with high reliability and bounded delay.
While effective, it hijacks cells allocated to existing traffic flows when an emergency is detected.
In the realm of reinforcement learning, Bommisetty et al.~\cite{bommisetty2022resource} introduce a Phasic Policy Gradient (PPG) based \acrshort{tsch} schedule learning algorithm.
This algorithm formulates the scheduling policy as an optimization problem, outperforming totally distributed and totally centralized \acrshort{drl}-based scheduling algorithms through the use of the actor-critic policy gradient method.

In summary, \acrshort{hrltsch} distinguishes itself from the surveyed approaches in several key aspects, as illustrated in Table~\ref{tab:comparison} that compares the surveyed approaches based on their architecture, \acrshort{ml} approach, dynamic requirements support, dynamic schedule support, optimization target, and key features.
\begin{enumerate}
    \item \textit{Focus on Schedule Design:} While previous works, such as~\cite{hajizadeh2023decentralized, jurado2023elise}, primarily concentrate on the parameter configuration of the \acrshort{tsch} protocol, \acrshort{hrltsch} takes a unique approach by placing its emphasis on the actual design of the \acrshort{tsch} schedule.
    \item \textit{Dual Policies:} In contrast to surveyed approaches~\cite{jurado2023elise,nguyen2019rl,dakdouk2018reinforcement,hajizadeh2023decentralized} that predominantly use a single policy, \acrshort{hrltsch} introduces a dual-policy framework.
          This dual-policy approach enhances the adaptability and optimization capabilities of the \acrshort{tsch} schedule.
    \item \textit{Awareness of Application Changes \highlight{(adaptability)}:} \acrshort{hrltsch} is designed to be aware of changes in application requirements, dynamically \highlight{adjusting} the \acrshort{tsch} schedule \highlight{in real-time} to maximize network performance based on evolving needs.
          This \highlight{adaptability is facilitated through two key mechanisms: Firstly, by adopting an \acrshort{sdn}-based architecture, HRL-TSCH enables swift, real-time modifications to the underlying \acrshort{tsch} schedule as necessitated by shifting user demands.
              Secondly, through the utilization of \acrshort{rl} in the policy formulation, the \acrshort{hrl} agent continuously seeks the optimal \acrshort{tsch} schedule based on current user requirements, thus enabling adaptive optimization in response to changing conditions.
              This capability sets HRL-TSCH apart from surveyed approaches, which generally lack awareness of application changes and dynamic adaptation, except for~\cite{jurado2023elise}, which optimizes the slotframe size.
          }
\end{enumerate}
To the best of our knowledge, \acrshort{hrltsch} marks the first attempt to introduce an \acrshort{hrl} approach dedicated to optimizing \acrshort{tsch} performance through tailored schedule design, tailored to meet the unique requirements of each application scenario.
For readers interested in a more comprehensive survey of \acrshort{tsch} schedulers, a detailed overview can be found in~\cite{urke2021survey, jerbi2022msu}.

\section{System Design}\label{sec:methodology}

\begin{table}[!htbp]
    \centering
    \caption{Notation}
    \label{tab:notation}
    \def\arraystretch{1}%
    \begin{NiceTabular}[c]{lm{6cm}}[vlines]
        \CodeBefore
        \rowcolor{lightgray}{1}
        \rowcolors{2}{gray!12}{}[respect-blocks]
        \Body
        \toprule
        \RowStyle[]{\bfseries}
        \textbf{Symbol}                                & \textbf{Description}                                                                                                                                    \\
        \midrule
        $\mathcal{N}$                                  & Set of nodes in the network                                                                                                                             \\
        $\mathcal{E}$                                  & Set of links between nodes                                                                                                                              \\
        $\lvert \mathcal{E} \lvert$                    & Number of links in the network                                                                                                                          \\
        $\mathcal{W}$                                  & Set of forwarding paths                                                                                                                                 \\
        $\mathcal{H}$                                  & Set of \acrshort{tsch} slots in the schedule                                                                                                            \\
        $\mathcal{U}$                                  & Set of timeslots in a \acrshort{tsch} schedule                                                                                                          \\
        $\lvert \mathcal{U} \lvert$                    & Number of timeslots in a \acrshort{tsch} schedule                                                                                                       \\
        $\lvert \mathcal{U}_{n,tx} \lvert$             & Number of transmitting timeslots of node $n$                                                                                                            \\
        $\mathcal{Z}$                                  & Set of channels in a \acrshort{tsch} schedule                                                                                                           \\
        $\lvert \mathcal{Z} \lvert$                    & Number of channels in a \acrshort{tsch} schedule                                                                                                        \\
        $F_{n,m}$                                      & Set of forwarding nodes of node $n$ in the forwarding path $m$                                                                                          \\
        $\varphi$                                      & Set of application requirements                                                                                                                         \\
        $\lvert u \rvert$                              & Duration of a timeslot                                                                                                                                  \\
        $T$                                            & Throughput of the network                                                                                                                               \\
        $T^{max}_{n,\lvert \mathcal{U}_{n,tx} \lvert}$ & Maximum throughput achieved by node $n$ with $\lvert \mathcal{U}_{n,tx} \rvert$ transmitting timeslots                                                  \\
        $T_0$                                          & Traffic in packets per second                                                                                                                           \\
        $T_{children,n}$                               & Incoming traffic from the children of node $n$                                                                                                          \\
        $B_n$                                          & Total traffic generated by node $n$                                                                                                                     \\
        $\xi_{T,n}$                                    & Noise term that accounts for the uncertainty in the throughput model of node $n$                                                                        \\
        $P$                                            & Power consumption of the network                                                                                                                        \\
        $P_n$                                          & Power consumption of node $n$                                                                                                                           \\
        $P_{n,tx}$                                     & Power consumption of node $n$ in the transmitting state                                                                                                 \\
        $P_{n,rx}$                                     & Power consumption of node $n$ in the receiving state                                                                                                    \\
        $E^f_{tx}$                                     & Energy consumption of the node $n$ in the transmitting state                                                                                            \\
        $E^f_{rx}$                                     & Energy consumption of the node $n$ in the receiving state                                                                                               \\
        $E^f_{rx\_ack}$                                & Energy consumption of the node $n$ in the receiving acknowledgment state                                                                                \\
        $E^f_{tx\_ack}$                                & Energy consumption of the node $n$ in the transmitting acknowledgment state                                                                             \\
        $E^f_{listen}$                                 & Energy consumption of the node $n$ in the idle listening state                                                                                          \\
        $\lvert \mathcal{H}^{idle}_{rx,n} \rvert$      & Number of cells in the \acrshort{tsch} schedule of node $n$ over a time span of one second that are in the receiving state, but no packets are received \\
        $P_0$                                          & Power consumption of \acrshortpl{sn} for basic operations                                                                                               \\
        $\xi_{p,n}$                                    & Noise term that accounts for the uncertainty in the power consumption model of node $n$                                                                 \\
        $D$                                            & Worst-case delay of the network                                                                                                                         \\
        $D(n, f, u)$                                   & Number of timeslots required for a data packet to travel from source node $n$ to forwarding node $f$ using timeslot $u$                                 \\
        $D(f, m)$                                      & Number of timeslots needed for a data packet to traverse from forwarding node $f$ to the next forwarding hop $f_m$ or the destination node              \\
        $D_{f,\mathcal{Q}}$                            & Delay of packets in the queue of forwarding node $f$                                                                                                    \\
        $\lambda_f$                                    & Arrival rate of packets at forwarding node $f$                                                                                                          \\
        $\mu_f$                                        & Service rate of packets at forwarding node $f$                                                                                                          \\
        $K$                                            & Large constant that represents the delay of packets in the queue of a forwarding node when the system is unstable                                       \\
        $\xi_{d,n}$                                    & Noise term that accounts for the uncertainty in the worst-case delay model of node $n$                                                                  \\
        $L$                                            & Loss rate of the network                                                                                                                                \\
        $\mathcal{R}$                                  & Set of rewards                                                                                                                                          \\
        $\mathcal{S}$                                  & Set of states                                                                                                                                           \\
        $\mathcal{A}$                                  & Set of actions                                                                                                                                          \\
        $\pi_{h}$                                      & Higher-level policy                                                                                                                                     \\
        $\pi_{l}$                                      & Lower-level policy                                                                                                                                      \\
        $\psi_h$                                       & Penalty for the higher-level policy                                                                                                                     \\
        $\psi_{\pi_l}$                                 & Penalty for the lower-level policy                                                                                                                      \\
        \bottomrule
    \end{NiceTabular}
\end{table}

This section presents an overview of the methods and techniques employed in this study to address the TSCH scheduling problem in \acrshort{iiot}.
By delving into the intricacies of the methodology, readers will gain a deeper understanding of how the research tackles the problem at hand and
how the proposed solution is designed and evaluated.
It also serves as a roadmap for understanding the core components and processes involved in
the research, providing insight into the system architecture, network model, and mathematical formulation of performance metrics.
The notation used throughout the paper is summarized in Table~\ref{tab:notation}.

\subsection{System Architecture}

\begin{figure}[t]
    \centering
    \includegraphics[width=\columnwidth]{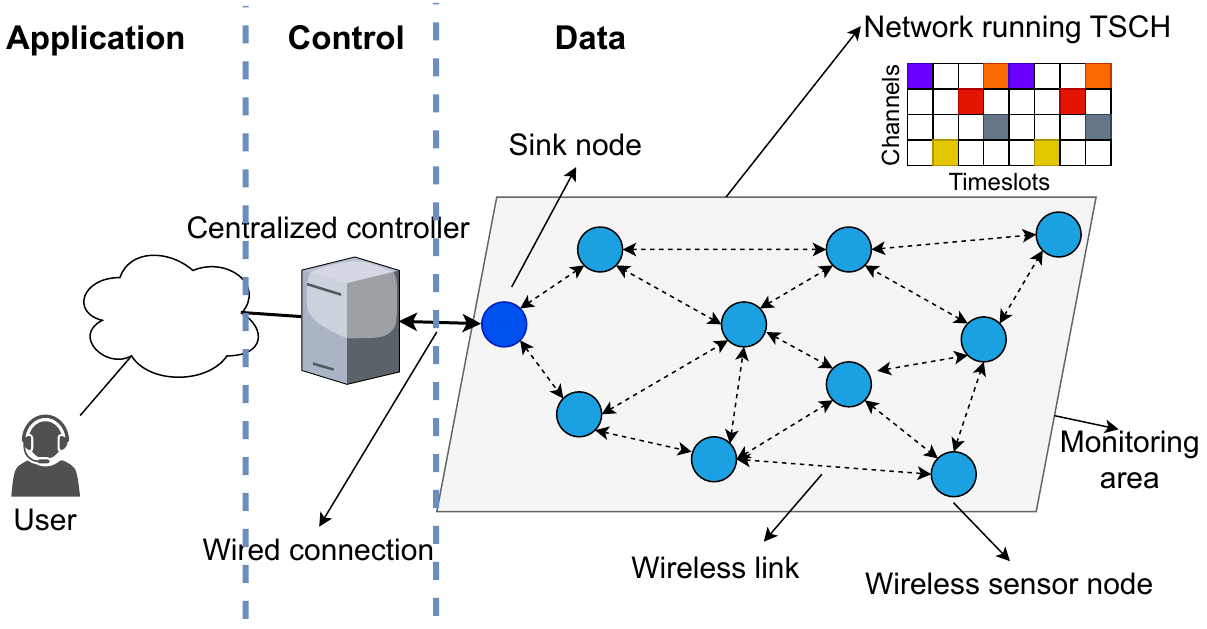}
    \caption{System architecture.}
    \label{fig:system_architecture}
\end{figure}

In this paper, we adopted the three-tier network architecture principles proposed in~\cite{jurado2023elise} to design a \acrshort{rl}-based scheduler that is aware of changes in the application
requirements and adapts the \acrshort{tsch} schedule accordingly to maximize the network performance.
The overall architecture is shown in Fig.~\ref{fig:system_architecture} and consists
of three planes: the application, control, and data plane.
\textit{The application plane} is the entrance point of the system and is responsible for receiving the weights of the application requirements from the user and sending them to the control plane.
\textit{The control plane} hosts multiple modules for the correct operation of the system including
the data collection and the network management module~\cite{shiny2022control}.
The network manager module provides an \acrfull{api} to reconfigure the network.
The control plane also hosts modules for the generation of forwarding paths and \acrshort{tsch} schedules, and the \acrshort{rl} agent.
Lastly, \textit{the data plane} is the network infrastructure built upon \acrshortpl{sn} (light blue nodes in Fig.~\ref{fig:system_architecture}) that collect the data of physical
phenomena of interest from the \textit{monitoring area} (e.g., industrial plant, agricultural field, etc.) and enforces, at each node, the forwarding paths and the \acrshort{tsch} communication links generated by the \acrshort{rl} agent and distributed by the control plane.
\acrshortpl{sn}, in the data plane, exchange packets with the control plane to report their status and receive the new forwarding paths and \acrshort{tsch} schedules using the \textit{sink node} (dark blue node in Fig.~\ref{fig:system_architecture}) as a gateway in a multi-hop fashion.
For a more comprehensive understanding of each plane and its modules, refer to~\cite{jurado2023elise}.

Given that the training of the \acrshort{rl} agent occurs offline, constructing a surrogate environment becomes imperative to emulate the characteristics of a real-world \acrshort{tsch} network.
Notably, online training within a testbed or network simulator is deemed impractical due to potential slowdown in the training process.
In the upcoming section, we delve into the network model and present the mathematical formulation of the performance metrics employed in this study.

\subsection{TSCH Network Model}

We model the network as a graph $G = (\mathcal{N}, \mathcal{E})$, where $\mathcal{N}$ is the set of nodes in the network, and $\mathcal{E}$ is the set of communication links between nodes.
Each node $n \in \mathcal{N}$ is associated with a set of attributes, such as the node's position, forwarding paths, \acrshort{tsch} schedule, and traffic load.
They can communicate directly with neighboring nodes within its transmission range.
Thus, we have defined the set of communication links within the transmission range as:
\begin{equation}
    \begin{split}
        \mathcal{E} = \{ &(n_x, n_y) \mid n_x, n_y \in N, x \neq y \text{ and node } n_y \\
        &\text{ is within the transmission range of node } n_x \}
    \end{split}
\end{equation}
\highlight{In our network model, each node can act as a source, a destination, or a relay node, except for the sink node.
    A source node generates data packets and sends them toward a destination node, while a destination node receives data packets from a source node.
    Additionally, a relay node facilitates data transmission by forwarding packets from the source node to the destination node.
}

The centralized entity, which is later introduced, that manages the network in this study is called the \textit{controller}.
The controller sets the forwarding paths $w \in \mathcal{W}$ and the \acrshort{tsch} slots $h \in \mathcal{H}$ of the network at any given time.
The forwarding paths are the paths that data packets follow from a source node to a destination node.
The set $\mathcal{H}$ is the \acrshort{tsch} schedule that determines the timeslots $u \in \mathcal{U}$ and channel offsets $\zeta \in \mathcal{Z}$ that nodes use to transmit and receive data packets.

\subsection{Mathematical Formulation of The Performance Metrics}
This section presents the mathematical formulation of the performance metrics to optimize in the \acrshort{tsch} scheduler.
These metrics also serve as a benchmark for assessing the system's capabilities and comparing it with other scheduling approaches.

\subsubsection{Throughput}
We use the throughput of the network as a measure of the efficiency of the network.
We express the throughput of the network ($T$) as a function of the number of packets delivered to the controller over a period of time.
The maximum throughput, in packets per second, achieved by node $n$ given that it has $\lvert \mathcal{U}_{n,tx} \lvert$ transmitting timeslots is calculated as follows:
\begin{equation}
    T^{max}_{n,\lvert \mathcal{U}_{n,tx} \rvert} = \frac{\lvert \mathcal{U}_{n,tx} \lvert}{\lvert \mathcal{U} \rvert \times \lvert u \rvert}
\end{equation}
Where $T^{max}_{n,\lvert \mathcal{U}{n,Tx} \lvert}$ represents the maximum throughput achieved by node $n$ with $\lvert \mathcal{U}_{n,tx} \rvert$ transmitting timeslots.
Then, the throughput of the network can be expressed as follows:
\begin{equation}\label{eq:throughput}
    T_n =
    \begin{cases}
        B_n,                                          & \text{if } T_{children,n}<T^{max}_{n,\lvert \mathcal{U}_{n,tx} \lvert} - T_0 \\
        T^{max}_{n,\lvert \mathcal{U}_{n,tx} \lvert}, & \text{otherwise}
    \end{cases}
\end{equation}
Where $T_0$ is the traffic in packets per second that is generated by each $n \in \mathcal{N}$, this includes control and data packets.
$T_{children,n}=\sum_{c \in \mathcal{N}_{child,n}} T_c$, depicts the incoming traffic from the children of node $n$, and $B_n = T_0 + T_{children,n}$ is the total traffic generated by node $n$.
The throughput of the network is then calculated as follows:
\begin{align}\label{eq:throughput_network}
    T                       & = \frac{\sum\limits_{n \in \mathcal{N}} \left( T_n + \xi_{T,n}\right)}{\lvert \mathcal{N} \lvert}                  \\
    \text{subject to} \quad & \lvert \mathcal{U} \lvert > 0,~\lvert \mathcal{N} \lvert > 0, \lvert u \lvert > 0 \label{eq:throughput_constraint}
\end{align}
Where $\xi_{T,n} \sim N(0, \sigma_{T,n}^2)$ is a noise term that accounts for the uncertainty in the throughput model of node $n$.

\subsubsection{Power Consumption}
In \acrshortpl{sn}, the power consumption ($P$) is significantly influenced by different radio communication states, such as transmit and receive states~\cite{dunkels2011powertrace}.
However, in TSCH networks, these states do not have equal contributions to node power consumption.
Receiving timeslots is likely to contribute more to the power consumption since \acrshortpl{sn} activate their radios even when there are no packets to receive.
On the other hand, transmitting timeslots is likely to contribute less to the power consumption since \acrshortpl{sn} only activate their radios when there are packets to transmit~\cite{vilajosana2013realistic,scanzio2020wireless}.
We assume that the power consumption of the network is mainly attributed to the receiving and transmitting states.
We can then calculate the power consumption of node $n$ for both states as follows:
\begin{align}
    P_{n,tx} & =T_n \times (E^f_{tx}+E^f_{rx\_ack})                                                                          \\
    P_{n,rx} & =T_{children,n} \times (E^f_{rx}+E^f_{tx\_ack}) + \lvert \mathcal{H}^{idle}_{rx,n} \rvert \times E^f_{listen}
\end{align}
Where $E^f_{tx}$, $E^f_{rx}$, $E^f_{rx\_ack}$, $E^f_{tx\_ack}$, and $E^f_{listen}$ are the energy consumption of the node $n$ in the transmitting, receiving, receiving acknowledgment, transmitting acknowledgment, and idle listening states, respectively.
$\lvert \mathcal{H}^{idle}_{rx,n} \rvert = T^{max}_{n,\lvert \mathcal{U}_{n,Rx} \rvert} - T_{children,n}$ is the number of cells in the \acrshort{tsch} schedule of node $n$ over a time span of one second that are in the receiving state, but no packets are received.
Then, $P_n = P_0 + P_{n,tx} + P_{n,rx} + \xi_{p,n}$, where $P_0$ captures the power consumption of \acrshortpl{sn} for basic operations, such as neighbor discovery and synchronization, and $\xi_{p,n} \sim N(0, \sigma_{p,n}^2)$ is a noise term that accounts for the uncertainty in the power consumption model of node $n$.
Therefore, the network power consumption can be expressed as follows:
\begin{align}\label{eq:power_consumption_network}
    P                       & = \frac{\sum\limits_{n \in \mathcal{N}} \left( P_n + \xi_{p,n}\right)}{\lvert \mathcal{N} \rvert} \\
    \text{subject to} \quad & \begin{aligned}
                                   & \lvert \mathcal{N} \rvert > 0, E^f_{tx} > 0, E^f_{rx} > 0, E^f_{rx\_ack} > 0, \\
                                   & E^f_{tx\_ack} > 0, E^f_{listen} > 0
                              \end{aligned} \label{eq:power_consumption_constraint}
\end{align}
\subsubsection{Worst-case Delay}
Our objective is to minimize the worst-case delay ($D$) in the network, which refers to the delay of the data packet that experiences the longest delivery time.
We use the notation $D(n, f, u)$ to denote the number of timeslots required for a data packet to travel from source node $n$ to forwarding node $f \in F_{n,m} \subset \mathcal{W}$ using timeslot $u \in \mathcal{U}$.
Similarly, $D(f, m)$ represents the number of timeslots needed for a data packet to traverse from forwarding node $f$ to the next forwarding hop $f_m$ or the destination node.
It's important to note that forwarding nodes relay data packets using their closest scheduled link to the next hop.
Mathematically, we can express the delay of node $n$ as follows:
\begin{equation}\label{eq:delay_node}
    D_n = \max\limits_{u \in \mathcal{U}} \left( D(n,f,u) + \sum\limits_{f \in F} D(f,f_m) \right)\times \lvert u \rvert + \sum\limits_{f \in F} D_{f,\mathcal{Q}}
\end{equation}
Where $\lvert u \lvert$ represents the duration of a timeslot.
$D_{f,\mathcal{Q}}$ is the delay, in milliseconds, of packets in the queue of forwarding node $f$, and it is calculated as follows:
\begin{equation}\label{eq:delay_queue}
    D_{f,\mathcal{Q}}=
    \begin{cases}
        \frac{\lambda_f}{\mu_f \times (\mu_f-\lambda_f)} \times 10^3, & \text{if } \lambda_f < \mu_f \\
        K,                                                            & \text{otherwise}
    \end{cases}
\end{equation}
Where $K$ is a large constant that represents the delay of packets in the queue of forwarding node $f$ when the system is unstable, meaning that the arrival rate of packets is higher than the service rate causing the queue to grow indefinitely.
$\lambda_f=T_{children,f}$ and $\mu_f=T_f$ are the arrival rate and service rate of packets at forwarding node $f$, respectively.
Therefore, the delay of the network can be expressed as follows:
\begin{align}\label{eq:delay_network}
    D                       & = \frac{\sum\limits_{n \in \mathcal{N}} \left( D_n + \xi_{d,n}\right)}{\lvert \mathcal{N} \lvert} \\
    \text{subject to} \quad & \lvert \mathcal{N} \lvert > 0, \lambda_f /\mu_f< 1, K >> 0 \label{eq:delay_constraint}
\end{align}
Where $\xi_{d,n} \sim N(0, \sigma_{d,n}^2)$ is a noise term that accounts for the uncertainty in the delay model of node $n$.

\section{Reinforcement Learning}\label{sec:reinforcement_learning}

This section provides an overview of the \acrshort{hrl} framework.
It also discusses the reward function used to guide the learning process of the \acrshort{rl} agents.
Lastly, we discuss the action space and the \acrshort{rl} algorithm used for training the agent.

\subsection{Reinforcement Learning Overview}
Designing an optimal \acrshort{tsch} schedule, which determines timeslots, channels, and slotframe size for efficient packet transmission and reception, is a challenging combinatorial optimization problem.

To tackle the complexity of the link assignments problem in a \acrshort{tsch} network, which intensifies with network size, timeslots and channels, and application requirements, we adopt a \acrfull{hrl} approach.
This approach empowers us to solve the problem effectively through a trial-and-error learning process, where the agent discovers the optimal mapping between state $s \in \mathcal{S}$ and action $a \in \mathcal{A}$ that maximizes the cumulative reward $\mathcal{R}$.

\begin{figure}[t]
    \centering
    \includegraphics[width=\columnwidth]{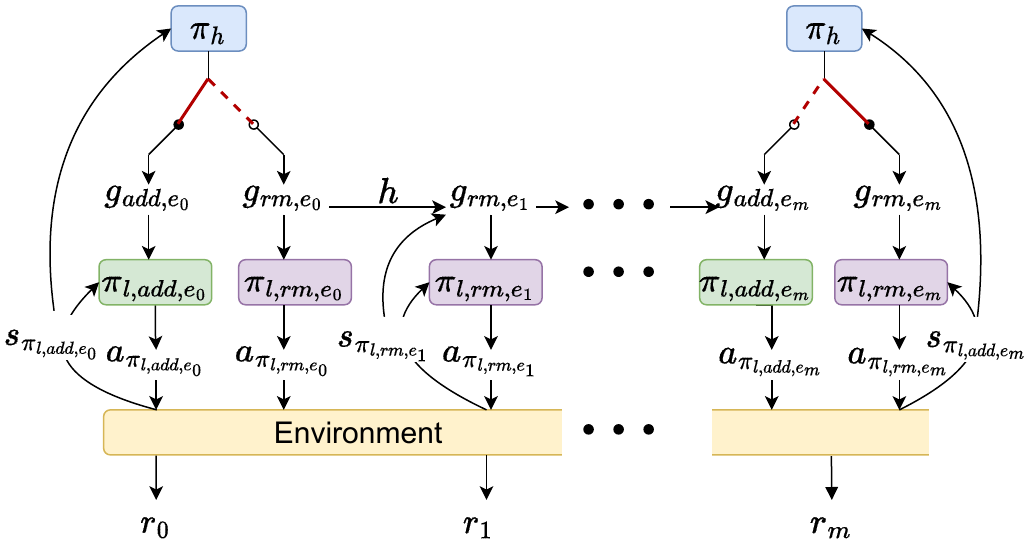}
    \caption{Reinforcement learning architecture.}
    \label{fig:rl_architecture}
\end{figure}

\subsection{Hierarchical Reinforcement Learning Architecture}
We extend the \acrshort{rl} framework to a \acrshort{hrl} framework to solve the link assignments problem in a \acrshort{tsch} network with one higher-level policy $\pi_{h}$ and multiple lower-level policies $\pi_{l,e}$ as shown in Fig.~\ref{fig:rl_architecture}.
Below we discuss the cost function used to guide the learning process of the \acrshort{rl} agents.

\subsubsection{Cost Function}
The cost function is a function that maps the state \( s \in \mathcal{S} \) and the action \( a \in \mathcal{A} \) to a cost \( c \in \mathcal{C} \), i.e., \( c: \mathcal{S} \times \mathcal{A} \to \mathbb{R} \).
Let $\varphi = (\alpha, \beta, \gamma)$ be the tuple of application requirements, where $\alpha$, $\beta$, and $\gamma$ are the weights of the normalized power consumption ($\widehat{P}$), delay ($\widehat{D}$), and throughput ($\widehat{T}$) respectively.
Therefore, the cost function can be expressed as follows:
\begin{align}\label{eq:cost_function}
     & c(s, a) = \alpha \times \widehat{P_t} + \beta \times \widehat{D_t} - \gamma \times \widehat{T_t} \\
     & \text{subject to} \quad \alpha + \beta + \gamma = 1 \label{eq:cost_function_constraint}
\end{align}
The values of $\varphi$ are provided by the application layer when the \acrshort{rl} agent is doing the inference.
During the learning process, the environment generates random $\varphi$ to train the \acrshort{rl} agent.
Maximum performance is achieved when the cost $c$ is minimized as shown below:
\begin{equation}\label{eq:min_cost_function}
    \begin{aligned}
        \min_{a \in \mathcal{A}} \quad & c(s, a)                                                                                                                                     \\
        \text{subject to} \quad        & (\ref{eq:throughput_constraint}), (\ref{eq:power_consumption_constraint}), (\ref{eq:delay_constraint}), (\ref{eq:cost_function_constraint})
    \end{aligned}
\end{equation}
We have now introduced the cost function used to guide the learning process of the \acrshort{rl} agent.
We will now focus on the two levels of the \acrshort{hrl} architecture.

\subsubsection{Higher-Level Policy}
The higher-level policy $\pi_{h}$ is responsible for selecting the optimal lower-level policy $\pi_{l,e}$, where $\pi_{l,e}$ is associated with a specific communication link $e \in \mathcal{E}$.
In brief, $\pi_{h}$ is in charge of selecting between two main actions:
\begin{enumerate}
    \item \textit{Add link}: This action adds a link $e$ to the \acrshort{tsch} schedule.
    \item \textit{Remove link}: This action removes a link $e$ from the \acrshort{tsch} schedule.
\end{enumerate}
Thus, the action space $\mathcal{A}$ of the higher-level policy $\pi_{h}$ is defined as follows:
\begin{equation}\label{eq:action_space_h}
    \mathcal{A}_{\pi_h} = \{a_{\pi_h} \mid a_{\pi_h} \in \{\mathcal{A}_{\pi_{h},add}, \mathcal{A}_{\pi_{h},rm}\}\}
\end{equation}
The terms $\mathcal{A}_{\pi_{h},\text{add}}$ and $\mathcal{A}_{\pi_{h},\text{rm}}$ represent specific subsets of the action space $\mathcal{A}_{\pi_h}$. Recall that $\mathcal{E}$ is the set of all possible communication links. We can define $\mathcal{A}_{\pi_{h},\text{add}}$ as the set of elements $a_{\pi_{h},\text{add}}$ such that each element represents a combination of communication links $e$ chosen from $\mathcal{E}$.
Formally, $\mathcal{A}_{\pi_{h},\text{add}} = \{a_{\pi_{h},\text{add}} \mid \forall e \in \mathcal{E}, a_{\pi_{h},\text{add}} \text{ represents the combination of } e\}$.
Thus, the size of the subset $\mathcal{A}_{\pi_h,add}$ is $\lvert \mathcal{E} \lvert$.
Similarly, $\mathcal{A}_{\pi_{h},\text{rm}}$ can be defined as the set of elements $a_{\pi_{h},\text{rm}}$ such that each element represents a combination of communication links $e$ chosen from $\mathcal{E}$ for removal.
Formally, $\mathcal{A}_{\pi_{h},\text{rm}} = \{a_{\pi_{h},\text{rm}} \mid \forall e \in \mathcal{E}, a_{\pi_{h},\text{rm}} \text{ represents the removal of } e\}$.
The size of the subset $\mathcal{A}_{\pi_h,rm}$ is also $\lvert \mathcal{E} \lvert$.
Therefore, the size of the action space $\mathcal{A}_{\pi_h}$ is $2 \times \lvert \mathcal{E} \lvert$.
The state space $\mathcal{S}$ of the higher-level policy $\pi_{h}$ is defined as follows:
\begin{equation}\label{eq:state_space_h}
    \mathcal{S}_{\pi_h} = \{s_{\pi_h} = (\widehat{P}, \widehat{D}, \widehat{T}, \varphi, \mathcal{\widehat{W}}, \mathcal{\widehat{H}},\widehat{e})\}
\end{equation}
Where $\mathcal{\widehat{W}}$ is a normalized adjacency array of $\mathcal{N}^2$ elements representing the network topology.
$\mathcal{\widehat{H}}$ is a normalized array of $\lvert \mathcal{Z} \rvert \times \lvert \mathcal{U} \rvert$ elements representing the \acrshort{tsch} schedule.
$\widehat{e}$ is the normalized value of the link $e$.
$s_{\pi_h}$ is provided by the environment to the \acrshort{rl} agent at each time step $t$.

The reward function of policy $\pi_{h}$ is defined as a function of the action taken, the state $s_{\pi_h}$, and the cost function $c$.
To discourage the policy $\pi_{h}$ from selecting actions that lead to higher costs or actions that are not feasible, we define a set of penalized actions $\mathcal{P}_{\pi_h}$.
One type of penalized action for the policy $\pi_{h}$ is denoted as $\mathcal{P}_{\pi_{h,add}} \subset \mathcal{P}_{\pi_h}$.
This type of action corresponds to adding a link $e$ that does not exist in the current forwarding paths $w \in \mathcal{W}$.
Formally, $\mathcal{P}_{\pi_{h,add}} = \{ p_{\pi_{h,add}} \mid p_{\pi_{h,add}} \in \mathcal{A}_{\pi_{h},\text{add}}, \forall e \in p_{\pi_h}, \neg \exists w \in \mathcal{W}, e \in q \} $
The condition $p_{\pi_h} \in \mathcal{A}_{\pi_{h},\text{add}}$ ensures that the penalized action $p_{\pi_h}$ is one of the actions in the set $\mathcal{A}_{\pi_{h},\text{add}}$.
The condition $\forall e \in p_{\pi_h}$ ensures that all the links $e$ being added in the action are considered.
Finally, the condition $\neg \exists w \in \mathcal{W}, e \in q$ ensures that there is no forwarding path $q$ in the set $Q$ that contains any of the links $e$ being added in the penalized action.

Another type of penalized action for the policy $\pi_{h}$ is denoted as $\mathcal{P}_{\pi_{h,rm}} \subset \mathcal{P}_{\pi_h}$.
This type of action corresponds to removing a link $e$ such as removing the link $e$ would result in a forwarding path $w \in \mathcal{W}$ that does not contain any links.
Formally, $\mathcal{P}_{\pi_{h,rm}} = \{ p_{\pi_{h,rm}} \mid p_{\pi_{h,rm}} \in \mathcal{A}_{\pi_{h},\text{rm}}, \forall e \in p_{\pi_h}, \exists w \in \mathcal{W}, e \in q \} $.
The condition $\exists w \in \mathcal{W}, e \in q$ ensures that there is a forwarding path $q$ in the set $Q$ that contains any of the links $e$ being removed in the penalized action.
We can now define the set of penalized actions as $\mathcal{P}_{\pi_h}= \{ p_{\pi_{h}} \mid p_{\pi_{h}} \in \mathcal{P}_{\pi_{h,add}} \cup \mathcal{P}_{\pi_{h,rm}} \}$.

The policy $\pi_{h}$ receives a penalty $\psi_{\pi_h}$ when it selects a penalized action $p_{\pi_h} \in \mathcal{P}_{\pi_h}$, and the episode ends as the agent has reached a terminal state.
The agent also reaches a terminal state when it has reached the maximum number of steps $T_{\pi_h}$.
Consequently, the immediate reward $r$ of the policy $\pi_{h}$ is defined as follows:
\begin{equation}\label{eq:reward_function_h}
    r_{\pi_h}(s_{\pi_h}, a_{\pi_h}) =
    \begin{cases}
        \psi_{\pi_h},                         & \text{if } a_{\pi_h} \in \mathcal{P}_{\pi_h} \\
        \upsilon-c(s_{\pi_{h}}, a_{\pi_{h}}), & \text{otherwise}
    \end{cases}
\end{equation}
Where $\upsilon>c(s_{\pi_h}, a_{\pi_h})$.
The objective is to find a policy $\pi_{h}^{*} = \arg\max_{\pi_{h}} \mathcal{R} = \arg\max_{\pi_{h}} \sum_{t=0}^{T} \lambda^{t} r_{\pi_h}(s_{\pi_h}, a_{\pi_h})
$ that maximizes the cumulative reward $\mathcal{R}$ over a time horizon.
Where $\lambda$ is the discount factor that determines the importance of future rewards.

We have now introduced the higher-level policy $\pi_h$, its action space, state space, and reward function.
We will now discuss the lower-level policy $\pi_{l,e}$.

\subsubsection{Lower-level Policies}
As discussed earlier, the action space of the policy $\pi_h$ consists of $2 \times |\mathcal{E}|$ actions. To select the optimal cell $h= (u, \zeta)$, where $u \in \mathcal{U}$ and $\zeta \in \mathcal{Z}$, for each action $a_{\pi_h}$, we need to define a set of lower-level policies. Let $\mathcal{L}$ denote the set of lower-level policies.

The set $\mathcal{L}$ consists of $2 \times |\mathcal{E}|$ lower-level policies, denoted as $\pi_{l,e}$, where $e \in \mathcal{E}$.
Among these lower-level policies, $|\mathcal{E}|$ are dedicated to selecting the optimal slot $h$ for adding link $e$ to the \acrshort{tsch} schedule, while the remaining $|\mathcal{E}|$ policies are responsible for selecting the optimal slot $h$ to remove link $e$ from the \acrshort{tsch} schedule.
By defining the set $\mathcal{L}$, we can effectively handle the selection of optimal cells at the lower level within the hierarchical architecture.

Both types of lower-level policies share the same action space $\mathcal{A}_{\pi_{l,e}}$ and state space $\mathcal{S}_{\pi_{l,e}}$ for their respective link $e$.
The action space $\mathcal{A}$ of the lower-level policies $\pi_{l,e}$ is defined as $\mathcal{A}_{\pi_{l,e}} = \{ a_{\pi_{l,e}} \mid a_{\pi_{l,e}} \in \mathcal{H}\}$, where $\mathcal{H}$ is the set of all possible cells $h= (u, \zeta)$.
It is important to note that the action space $\mathcal{A}_{\pi{l,e}}$ is specific to each link $e$.
The number of actions in $\mathcal{A}_{\pi{l,e}}$ is equal to the number of cells in the \acrshort{tsch} schedule, which is given by $\lvert \mathcal{Z} \lvert \times \lvert \mathcal{U} \lvert$ for each link $e$.

The state space $\mathcal{S}$ of the lower-level policy $\pi_{l,e}$ for each link $e$ is defined as follows:
\begin{equation}\label{eq:state_space_l}
    \mathcal{S}_{\pi_{l,e}} = \{s_{\pi_{l,e}} = (\widehat{P}, \widehat{D}, \widehat{L}, \varphi, \mathcal{\widehat{W}}, \mathcal{\widehat{H}}\times k, \widehat{e}) \}
\end{equation}
Here, the state $s_{\pi_{l,e}}$ is provided by the environment to the RL agent at each time step $t$.
$\mathcal{\widehat{H}}\times k$ is an array that represents the TSCH schedule of the source and destination nodes and the network schedule occupation.
It is important to note that the state space $\mathcal{S}_{\pi_{l,e}}$ is specific to each link $e$.

The reward function of the two types of lower-level policies $\pi_{l,e}$ differs in the penalized set of actions $\mathcal{P}_{\pi_{l,e}}$.
We denote the set of occupied cells in state $s_{\pi_{l,e}}$ as $H(s_{\pi_{l,e}})$.
For the policy $\pi_{l,add}$, we define a set of penalized actions $\mathcal{P}_{l,add} = \{a_{\pi_{l,e}} \mid a_{\pi_{l,e}}= h= (u, \zeta),u \in \mathcal{U},\zeta \in \mathcal{Z}, h\in H(s_{\pi_{l,e}})\}$
to discourage the selection of infeasible or suboptimal actions.
The penalized actions include selecting cells that are already occupied in the current TSCH schedule, adding a transmission link to a cell already occupied by the source node, or adding a reception link to a cell already occupied by the destination node.
By including the penalized set $\mathcal{P}_{l,add}$ in the learning, we guide the agent towards policies that prioritize optimal cell selection while avoiding actions that may have negative consequences on system performance.
Besides, it promotes the exploration of more efficient and non-overlapping solutions.

For the policy $\pi_{l,rm}$, we define a set of penalized actions $\mathcal{P}_{l,\text{rm}} = \{a_{\pi_{l,e}} \mid a_{\pi_{l,e}}= h= (u, \zeta), u \in \mathcal{U}, \zeta \in \mathcal{Z}, h\notin H(s_{\pi_{l,e}}) \text{ or } m(h) \neq m(s_{\pi_{l,e}}(h))\}$.
Where $m(h)$ represents the destination node associated with slot $h$, and $m(s_{\pi_{l,e}}(h))$ represents the destination node associated with the scheduled link in cell $h$.
The condition $h\notin H(s_{\pi_{l,e}})$ ensures that the action selects a cell that is already occupied, preventing the agent from attempting to remove a nonexistent link from an empty cell.
The condition $m(h) \neq m(s_{\pi_{l,e}}(h))$ checks if the link in cell $h$ points to a different destination node.
If this condition is met, it indicates a mismatch between the scheduled link configuration and the agent's removal action, which could disrupt the established communication paths and lead to suboptimal performance.
By penalizing actions in $\mathcal{P}_{l,\text{rm}}$, we guide the agent to focus on removing redundant links only from valid cells while avoiding actions that could cause disruptions or inconsistencies in the network topology.
We can now define the set of penalized actions in the lower-level policy $\pi_{l,e}$ as $\mathcal{P}_{\pi_{l,e}} = \{ p_{\pi_{l,e}} \mid p_{\pi_{l,e}} \in \mathcal{P}_{l,add} \cup \mathcal{P}_{l,\text{rm}} \}$.

The immediate reward $r$ of the lower-level policy $\pi_{l,e}$ is defined as follows:
\begin{equation}\label{eq:reward_function_l}
    r_{\pi_{l,e}}(s_{\pi_{l,e}}, a_{\pi_{l,e}}) =
    \begin{cases}
        \begin{aligned}
             & \psi_{\pi_l}, \quad &  & \text{if } a_{\pi_{l,e}} \in \mathcal{P}_{\pi_{l,e}}
        \end{aligned} \\
        \begin{aligned}
             & \upsilon-c(s_{\pi_{l,e}}, a_{\pi_{l,e}}), &  & \text{otherwise}
        \end{aligned}
    \end{cases}
\end{equation}
The objective here is also to find a policy $\pi_{l,e}$ that maximizes the cumulative reward $\mathcal{R}$ over a time horizon, as in $\pi_{h}$.

\begin{algorithm}[htbp]
    \footnotesize %
    \caption{HRLTSCH Algorithm}
    \label{alg:hrltsch}
    \SetAlgoLined
    \SetKwComment{Comment}{\textcolor{blue}{/* }}{\textcolor{blue}{ */}}
    \SetKwInOut{Input}{Input}

    \Input{Replay memory capacity $\mathcal{D}$, batch size $\mathcal{B}$, target network update rate $\alpha_{\theta^{-}}$, discount factor $\lambda$, initial exploration rate $\epsilon$, minimum exploration rate $\epsilon_{\text{min}}$, and exploration rate decay $\epsilon_{\text{decay}}$}

    Notations: $\theta$ and $\theta^{-}$ are the weights of the Q-network and target network, respectively.

    Initialize replay memory $\mathcal{D}$, parameters $\theta$ and $\theta^{-}$ with random weights\;

    \For{Each episode}{
    Reset the environment ($\mathcal{H}$, $\varphi$)\;
    Observe the state $s_{\pi_h}$ from the environment\;

    \While{not done}{
    $\epsilon \gets \max(\epsilon_{\text{min}}, \epsilon_{\text{decay}} \times \epsilon)$\;
    Choose action $a_{\pi_h}$ using $\epsilon$-greedy policy\;
    \Comment{\textcolor{blue}{Select the optimal link $e$ to be added or removed}}
    \If{$a_{\pi_h} \in \mathcal{A}_{\pi_{h},\text{add}}$}{
    \Comment{\textcolor{blue}{Select the optimal cell for link $e$ to be added}}
    Choose action $a_{\pi_{l,e}}$ using $\epsilon$-greedy policy\;
    }
    \If{$a_{\pi_h} \in \mathcal{A}_{\pi_{h},\text{rm}}$}{
    \Comment{\textcolor{blue}{Select the optimal link $e$ to be removed}}
    Choose action $a_{\pi_{l,e}}$ using $\epsilon$-greedy policy\;
    }
    Execute action $a_{\pi_{l,e}}$, obtain reward $r_{\pi_{l,e}}$ using Eq. (\ref{eq:reward_function_l}) and next state $s'_{\pi_{l,e}}$\;
    Obtain reward $r_{\pi_h}$ using Eq. (\ref{eq:reward_function_h})\;
    Store transition $(s_{\pi_h}, a_{\pi_h}, r_{\pi_h}, s'_{\pi_h})$ in $\mathcal{D}$\;

    \If{$\lvert \mathcal{D} \rvert > \lvert \mathcal{B}\rvert $}{
    Sample a batch $(s, a, r, s')$ from $\mathcal{D}$\;
    \Comment{\textcolor{blue}{Calculate the loss and update the weights of the Q-network}}
    $L(\theta) \gets \frac{1}{\lvert \mathcal{B} \rvert} \sum_{i=1}^{\lvert \mathcal{B} \rvert} \left( y_{\pi_h} - Q(s_{\pi_h}, a_{\pi_h}; \theta) \right)^{2}$\;
    Update the weights of the Q-network\;
    \If{$\text{mod}(t, \alpha_{\theta^{-}}) = 0$}{
        Update the weights of the target network $\theta^{-} \gets \theta$\;
    }
    }
    }
    }
\end{algorithm}

\subsection{Training}

The \acrshort{hrltsch} algorithm undergoes training using the \acrshort{dqn} algorithm, as illustrated in Algorithm \ref{alg:hrltsch}.
It is important to note that the lower-level policies $\pi_{l,e}$ are trained independently of each other.
Once the lower-level policies $\pi_{l,e}$ are successfully trained, the higher-level policy $\pi_{h}$ is then trained.

The training process for the lower-level policies $\pi_{l,e}$ closely mirrors the training process for the higher-level policy $\pi_{h}$.
Following the training of the lower-level policies, the higher-level policy $\pi_{h}$ utilizes these trained policies to make informed decisions, selecting optimal actions $a_{\pi_{l,e}}$ for each link $e$.

We trained the \acrshort{rl} agents with $5\times 10^5$ steps using a replay memory capacity of $10^{5}$, a batch size of 512, a learning rate ($\sigma$) of 0.001, a learning start time of 5000, a discount factor ($\lambda$) of 0.8, an exploration fraction of 0.7, and a minimum exploration rate of 0.01.

\subsection{\acrshort{tsch} Lookup Algorithm}

Once the RL agents select the optimal links $e$ and cells $h$, the next step is to generate the \acrshort{tsch} schedule and distribute it to all $n \in \mathcal{N}$.
The \acrshort{tsch} schedule is a list of $e$, where each $e$ is associated with a source node, a destination node, a timeslot, and a channel offset.
The \acrshort{sdn} controller is responsible for translating the \acrshort{tsch} schedule into packets that are sent to the nodes in the network.
Each node $n \in N$ in the network receives the \acrshort{tsch} schedule and processes it to generate the \acrshort{tsch} schedule for its use.

\highlight{For the low-level policy, }Each $n \in N$ runs the Algorithm \ref{alg:tsch_link_selection} to determine the $u$ and $\zeta$ to use.
\highlight{
    This algorithm serves as a complementary mechanism to the RL-driven decision-making process.
    While RL agents determine the optimal links and cells to use, the TSCH lookup algorithm ensures that the selected links are correctly scheduled in the TSCH network.
}
It performs an iterative search through the list of $e$, comparing the destination address of each link with the provided destination address.
The algorithm calculates the difference between the $u$ of each $e$ and the current \acrfull{asn}, and it keeps track of the minimum difference found and stores the corresponding $u$ and $\zeta$.

\begin{algorithm}[t]
    \footnotesize %
    \caption[TSCH Link Selection]{TSCH Link Selection Algorithm}
    \label{alg:tsch_link_selection}
    \SetAlgoLined
    \SetKwComment{Comment}{\textcolor{blue}{/* }}{\textcolor{blue}{ */}}
    \SetKwInOut{Input}{Input}
    \SetKwInOut{Output}{Output}

    \SetKwFunction{gettsch}{get\_ts\_ch\_from\_dst\_addr}

    \SetKwProg{Fn}{Function}{}{end}

    \Fn{\gettsch{dst}}{
        \Input{Destination address \textit{dst}}
        \Output{$u$, $\zeta$}

        Initialize \textit{diff}, \textit{min} $\gets \infty$, $\lvert \mathcal{U} \lvert$ \;
        \Comment{\textcolor{blue}{Calculate the current time slot $u$}}
        $u_{ASN}\gets ASN\bmod \lvert \mathcal{U} \lvert$\;
        \Comment{\textcolor{blue}{Iterate through the list of links $e$}}
        $l \gets$ head of links $e$ list\;

        \While{ $l \neq \emptyset$}{
            \Comment{\textcolor{blue}{Check if the destination address of the link $e$ matches the provided destination address}}
            \If{$dst= l.dst$}{
                \Comment{\textcolor{blue}{Get the time slot $u$ with the minimum difference}}
                \textit{diff} $ \gets l.u - u_{ASN}$\;
                \If {\textit{diff}$ < $ 0}{
                    \Comment{\textcolor{blue}{We add to the difference the number of time slots $\lvert \mathcal{U} \lvert$ as the time slot $u$ is in the past}}
                    \textit{diff} $\mathrel{+}= \lvert \mathcal{U} \lvert$\;
                }
                \If{$\textit{diff} < $ \textit{min}}{
                    \Comment{\textcolor{blue}{Update the minimum difference, store the time slot $u$ and channel offset $\zeta$}}
                    \textit{min} $\gets$ \textit{diff} \;
                    $u \gets l.u$\;
                    $\zeta \gets l.\zeta$ \;
                }
            }
            $l \gets$ next link $e$ in the list\;
        }
        \Comment{\textcolor{blue}{Return the time slot $u$ and channel offset $\zeta$}}
        \Return $u$, $\zeta$\;
    }
\end{algorithm}

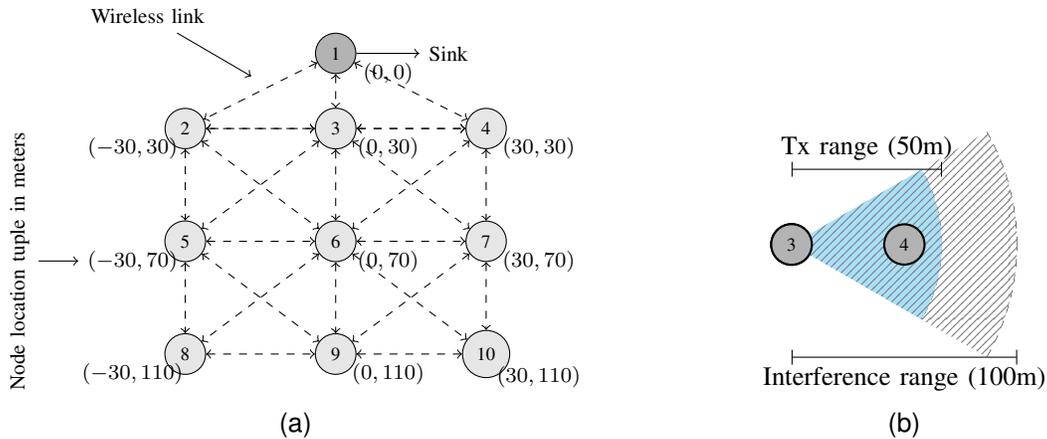
\begin{figure*}[t]
    \centering
    \subfloat[\label{fig:network_topology}]{
        \begin{tikzpicture}
            \centering
            \newlength{\circleradius}
            \setlength{\circleradius}{0.4cm}
            \def\rxoffset{0.7}
            \def\lxoffset{-0.7}
            \def\yoffset{-0.25}

            \node[circle, draw, fill=black!30, minimum size=\circleradius,font=\scriptsize] (sink) at (0, 0) {1};
            \node[anchor=south,font=\footnotesize] at ([shift={(\rxoffset, \yoffset)}]sink.south) {$(0,0)$};
            \node[circle, draw, fill=black!10, minimum size=\circleradius,font=\scriptsize] (sensor2) at (-2, -1) {2};
            \node[anchor=south,font=\footnotesize]  at ([shift={(\lxoffset, \yoffset)}]sensor2.south) {$(-30,30)$};
            \node[circle, draw, fill=black!10, minimum size=\circleradius,font=\scriptsize] (sensor3) at (0, -1) {3};
            \node[anchor=south,font=\footnotesize]  at ([shift={(\rxoffset, \yoffset)}]sensor3.south) {$(0,30)$};
            \node[circle, draw, fill=black!10, minimum size=\circleradius,font=\scriptsize] (sensor4) at (2, -1) {4};
            \node[anchor=south,font=\footnotesize]  at ([shift={(\rxoffset, \yoffset)}]sensor4.south) {$(30,30)$};
            \node[circle, draw, fill=black!10, minimum size=\circleradius,font=\scriptsize] (sensor5) at (-2, -2.5) {5};
            \node[anchor=south,font=\footnotesize]  at ([shift={(\lxoffset, \yoffset)}]sensor5.south) (pos_sensor5) {$(-30,70)$};
            \node[circle, draw, fill=black!10, minimum size=\circleradius,font=\scriptsize] (sensor6) at (0, -2.5) {6};
            \node[anchor=south,font=\footnotesize]  at ([shift={(\rxoffset, \yoffset)}]sensor6.south) {$(0,70)$};
            \node[circle, draw, fill=black!10, minimum size=\circleradius,font=\scriptsize] (sensor7) at (2, -2.5) {7};
            \node[anchor=south,font=\footnotesize]  at ([shift={(\rxoffset, \yoffset)}]sensor7.south) {$(30,70)$};
            \node[circle, draw, fill=black!10, minimum size=\circleradius,font=\scriptsize] (sensor8) at (-2, -4) {8};
            \node[anchor=south,font=\footnotesize]  at ([shift={(\lxoffset, \yoffset)}]sensor8.south) {$(-30,110)$};
            \node[circle, draw, fill=black!10, minimum size=\circleradius,font=\scriptsize] (sensor9) at (0, -4) {9};
            \node[anchor=south,font=\footnotesize]  at ([shift={(\rxoffset, \yoffset)}]sensor9.south) {$(0,110)$};
            \node[circle, draw, fill=black!10, minimum size=\circleradius,font=\scriptsize] (sensor10) at (2, -4) {10};
            \node[anchor=south,font=\footnotesize]  at ([shift={(\rxoffset, \yoffset)}]sensor10.south) {$(30,110)$};

            \draw[<->, dashed] (sink) -- (sensor3);
            \draw[<->, dashed] (sink) -- node[midway, above](node_midway){} (sensor2);
            \draw[<->, dashed] (sink) -- (sensor4);
            \draw[<->, dashed] (sensor3) -- (sensor2);
            \draw[<->, dashed] (sensor3) -- (sensor4);
            \draw[<->, dashed] (sensor2) -- (sensor5);
            \draw[<->, dashed] (sensor2) -- (sensor6);
            \draw[<->, dashed] (sensor2) -- (sensor3);
            \draw[<->, dashed] (sensor3) -- (sensor5);
            \draw[<->, dashed] (sensor3) -- (sensor6);
            \draw[<->, dashed] (sensor3) -- (sensor7);
            \draw[<->, dashed] (sensor3) -- (sensor4);
            \draw[<->, dashed] (sensor4) -- (sensor6);
            \draw[<->, dashed] (sensor4) -- (sensor7);
            \draw[<->, dashed] (sensor5) -- (sensor8);
            \draw[<->, dashed] (sensor5) -- (sensor9);
            \draw[<->, dashed] (sensor5) -- (sensor6);
            \draw[<->, dashed] (sensor6) -- (sensor8);
            \draw[<->, dashed] (sensor6) -- (sensor9);
            \draw[<->, dashed] (sensor6) -- (sensor10);
            \draw[<->, dashed] (sensor6) -- (sensor7);
            \draw[<->, dashed] (sensor7) -- (sensor9);
            \draw[<->, dashed] (sensor7) -- (sensor10);
            \draw[<->, dashed] (sensor8) -- (sensor9);
            \draw[<->, dashed] (sensor9) -- (sensor10);

            \node[align=right,thick,font=\footnotesize] (sink_label) at (1.5,0) {Sink};
            \draw[->] (sink) -- (sink_label);

            \node[align=left,thick,font=\footnotesize] (wireless_link_label) at (-2.5,0.5) {Wireless link};
            \draw[->] (wireless_link_label) -- (node_midway);

            \node[align=right,rotate=90,font=\footnotesize] (pos_sensor5_label) at ([shift={(-1.5,0)}]pos_sensor5) {Node location tuple in meters};
            \draw[->] (pos_sensor5_label) -- (pos_sensor5);

        \end{tikzpicture}
    }
    \hfil
    \subfloat[\label{fig:tx_if_range}]{
        \begin{tikzpicture}
            \centering
            \def\rxoffset{0.7}
            \def\lxoffset{-0.7}
            \def\yoffset{-0.25}
            \node[circle, draw, fill=black!30, thick, minimum size=0.4cm, font=\scriptsize] (sensor3) at (0, 0) {3};
            \node[circle, draw, fill=black!10, minimum size=0.4cm, font=\scriptsize] (sensor4) at (1.5, 0) {4};

            \def\innerRadius{20mm}
            \def\outerRadius{30mm}

            \begin{scope}
                \clip (sensor3.center) -- ($(sensor3.center) + (30:\innerRadius)$) arc (30:-30:\innerRadius) -- cycle;
                \draw[dashed] (sensor3.center) circle (\innerRadius);
                \fill[color=cyan!30] (sensor3.center) circle (\innerRadius);
            \end{scope}

            \begin{scope}
                \clip (sensor3.center) -- ($(sensor3.center) + (30:\outerRadius)$) arc (30:-30:\outerRadius) -- cycle;
                \draw[dashed] (sensor3.center) circle (\outerRadius);
                \fill[pattern=north east lines, pattern color=gray] (sensor3.center) circle (\outerRadius);
            \end{scope}

            \node[circle, draw, fill=black!30, thick, minimum size=0.4cm, font=\scriptsize] (sensor3) at (0, 0) {3};
            \node[circle, draw, fill=black!30, thick, minimum size=0.4cm, font=\scriptsize] (sensor4) at (1.5, 0) {4};

            \draw[|-|] (sensor3.center |- 0,1) -- node[above] {Tx range (50m)} (\innerRadius,1);
            \draw[|-|] (sensor3.center |- 0,-1.5) -- node[below] {Interference range (100m)} (\outerRadius,-1.5);

        \end{tikzpicture}
    }
    \caption{\highlight{Network topology and wireless link visualization. (a) Network topology with the sink node and sensor nodes. (b) Wireless link visualization showing the transmission range and interference range.}}
    \label{fig:network_topology_tx_range}
\end{figure*}

\section{Performance Evaluation}\label{sec:performance_evaluation}
We conducted our experiments in the Cooja simulator~\cite{osterlind2006cross} with retransmissions disabled to approximate the network closely to our model.
The simulations run on a single machine with an Intel Core i9 CPU with 16GB of \acrshort{ram}.
The network topology represents a small-scale network comprising ten \acrshortpl{sn}, as illustrated in Fig.~\ref{fig:network_topology}.
\highlight{
    Each \acrshort{sn} is 30 meters apart from its neighbors, and the sink node is located at the top of the network.
    All sensor nodes are equipped with an IEEE 802.15.4 radio transceiver and use the \acrfull{udgm} as the distance loss model.
    The transmission and interference range of the radio transceiver is set to 50 and 100 meters, respectively (see Fig.~\ref{fig:tx_if_range}).
    \acrshortpl{sn} are equipped with the Contiki-NG Energest module to calculate the energy consumption of \acrshortpl{sn} at each power state (e.g., transmit, receive, listen, and sleep).
}
In this configuration, the sink node is connected to the control plane via a serial interface, as depicted in Fig.~\ref{fig:system_architecture}.
We utilize this small-scale network to avoid excessive complexity in our experiments and to enable us to draw meaningful conclusions from the proof of concept of the proposed approach.
A summary of the network parameters is provided in Table~\ref{tab:network_parameters}.
\begin{table}[t]
    \centering
    \caption{Network parameters.}
    \label{tab:network_parameters}
    \def\arraystretch{1}%
    \footnotesize
    \begin{NiceTabular}{|l|l|l|l|}[cell-space-limits=1pt]
        \CodeBefore
        \rowcolor{lightgray}{1}
        \rowcolors{2}{gray!12}{}[respect-blocks]
        \Body
        \toprule
        \RowStyle[]{\bfseries}
        Parameter                                 & Value              & Parameter                     & Value                       \\
        \midrule
        $\mathcal{\lvert N \rvert}$               & $10$               & $E_{tx\_ack}$                 & $55$ $\mu$J                 \\
        $\mathcal{\lvert U \rvert}$               & $17$               & $E_{rx}$                      & $160$ $\mu$J                \\
        $\mathcal{\lvert Z \rvert}$               & $2$                & $E_{rx\_ack}$                 & $70$ $\mu$J                 \\
        $\lvert u \rvert$                         & $10$ ms            & $E_{listen}$                  & $110$ $\mu$J                \\
        $ \psi_{\pi_h}$,  $\psi_{\pi_l}$          & $-1$               & Operating voltage             & $3$ V                       \\
        $\upsilon$                                & $2$                & Tx current                    & $19.5$ mA                   \\
        $K$                                       & $10^3$ ms          & Rx current                    & $21.8$ mA                   \\
        $E_{tx}$                                  & $140$ $\mu$J       & CPU current                   & $1.85$ mA                   \\
        Deep LPM current                          & $0.0051$ mA        & LPM current                   & $0.0545$ mA                 \\
        \highlight{Dist. between \acrshortpl{sn}} & \highlight{$30$ m} & \highlight{Propagation model} & \highlight{\acrshort{udgm}} \\
        \highlight{Tx range}                      & \highlight{$50$ m} & \highlight{If range}          & \highlight{$100$ m}         \\
        \highlight{Data interval}                 & \highlight{$1$ s}  & \highlight{Packet size}       & \highlight{$12$ B}          \\
        \hline
    \end{NiceTabular}
\end{table}
\begin{figure*}[t]
    \centering
    \subfloat[\label{fig:pareto_front_power_consumption}]{\includegraphics[width=0.32\textwidth]{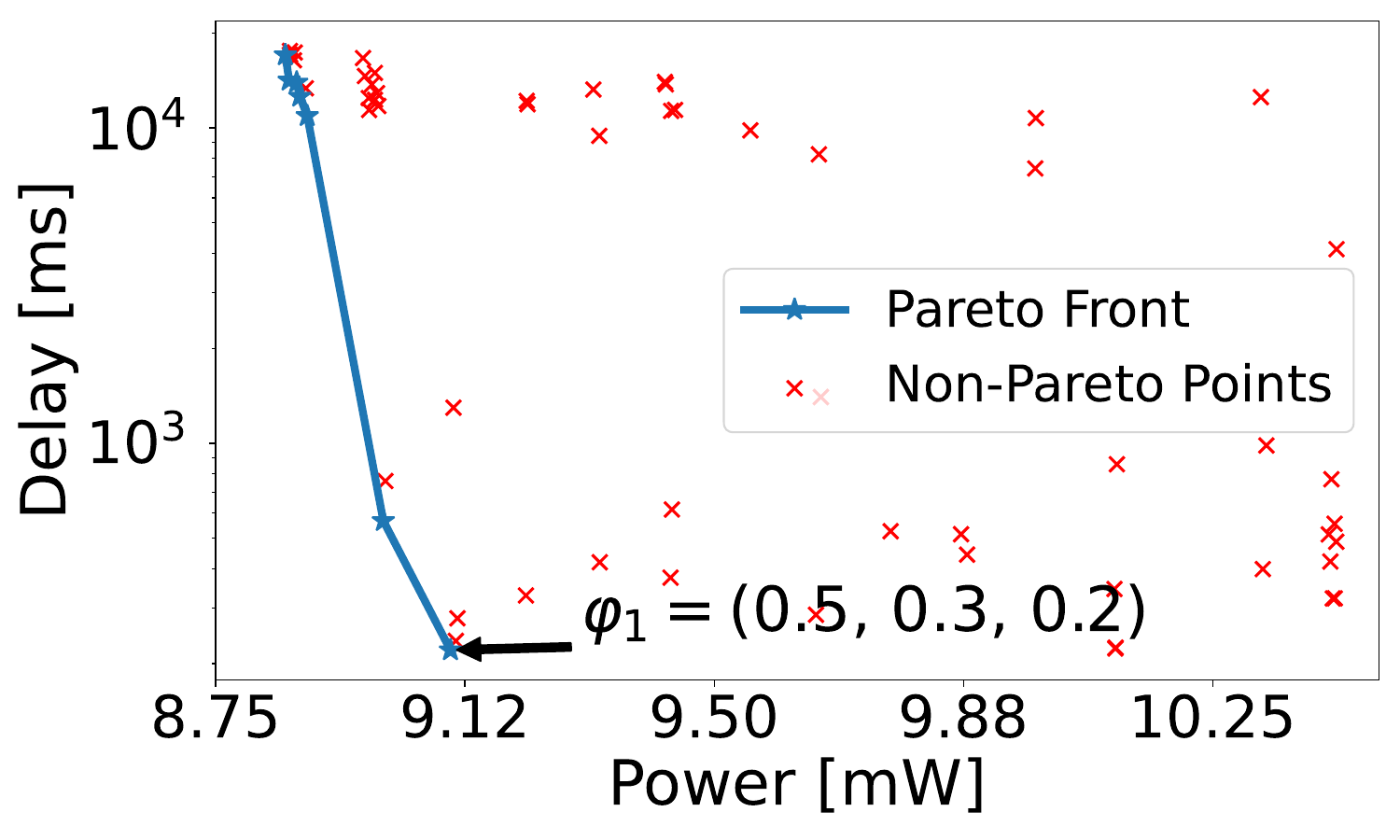}}\hfill
    \subfloat[\label{fig:pareto_front_latency}]{\includegraphics[width=0.32\textwidth]{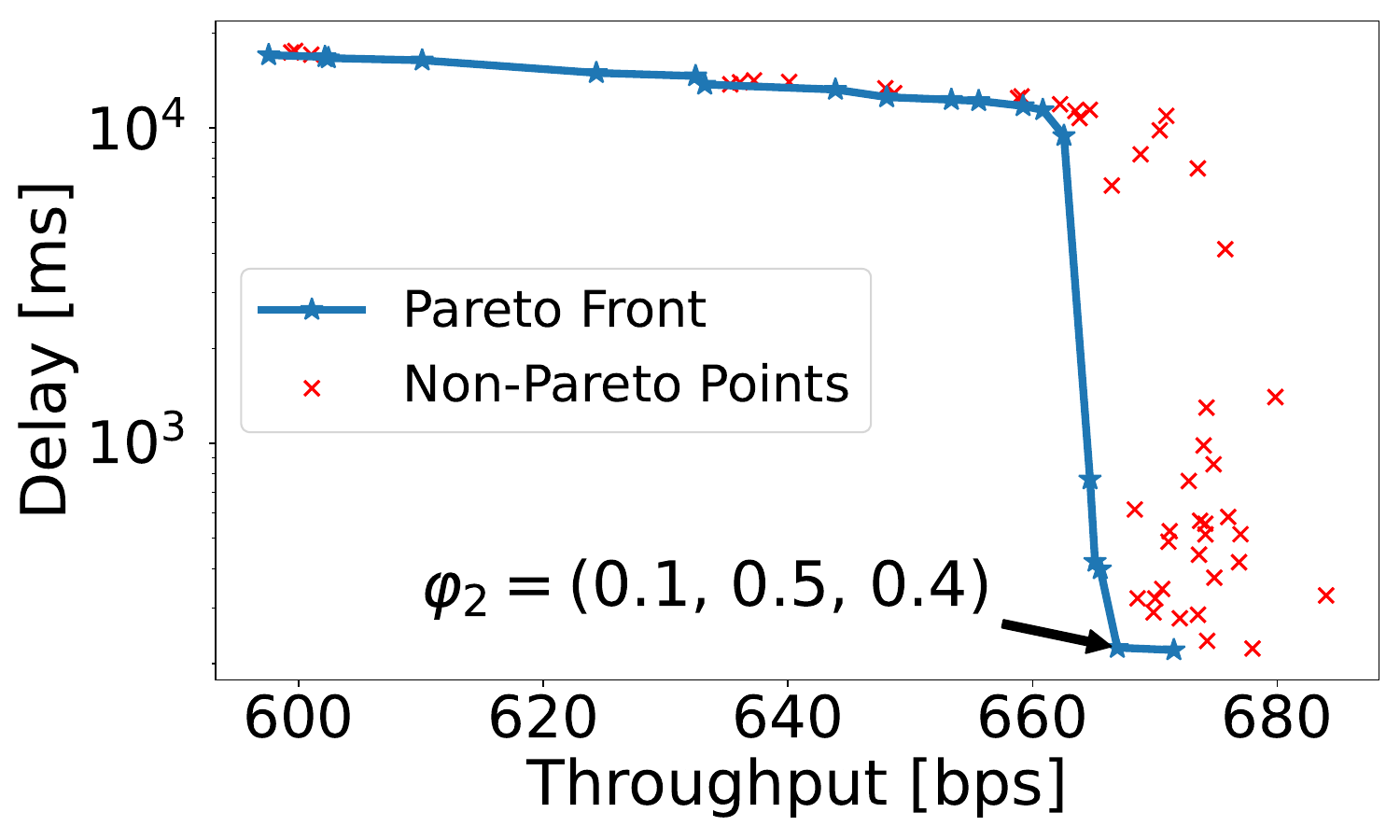}}\hfill
    \subfloat[\label{fig:pareto_front_throughput}]{\includegraphics[width=0.32\textwidth]{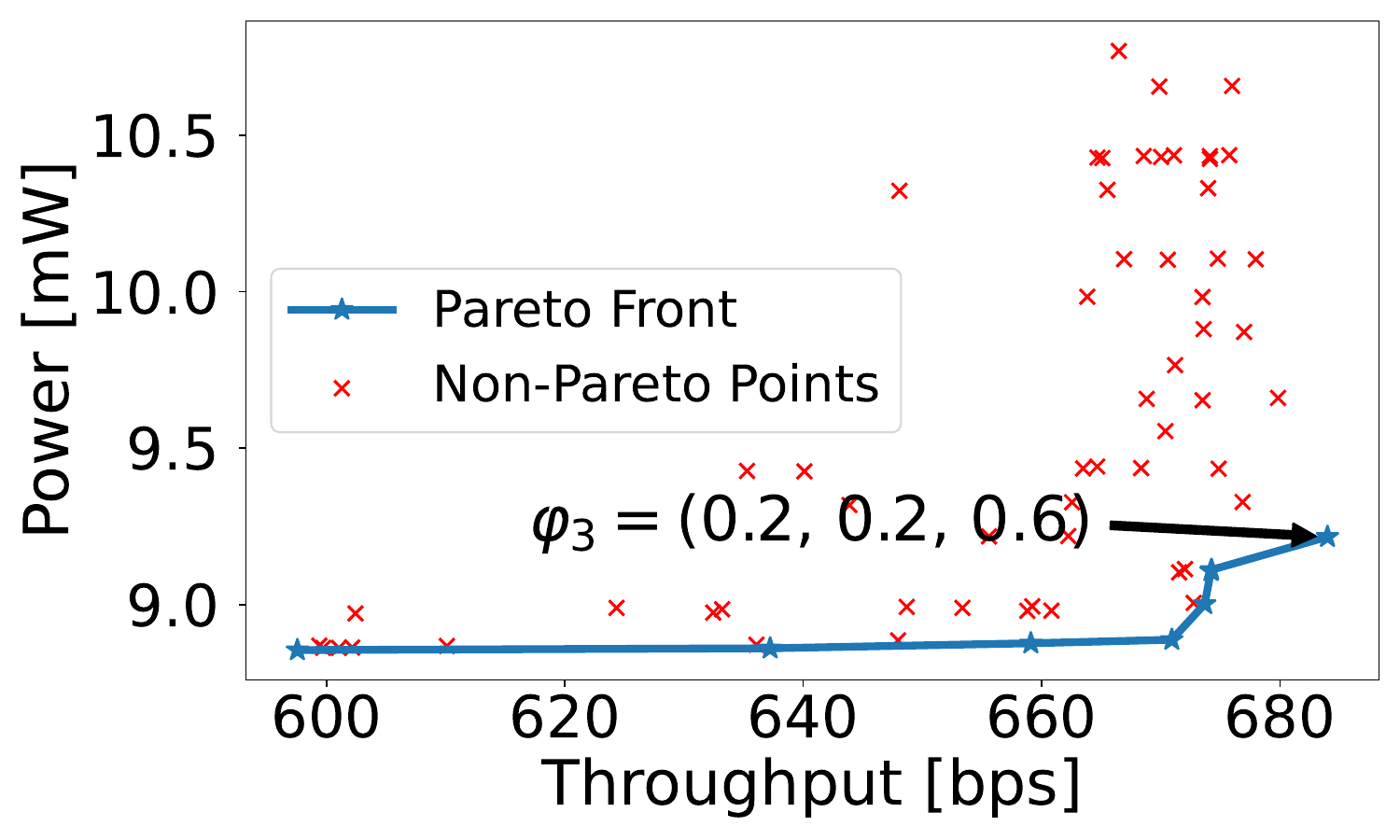}}
    \caption{Pareto front that shows the trade-offs between power consumption, delay, and throughput. Fig.~\ref{fig:pareto_front_power_consumption} shows the trade-offs between power consumption and delay. Fig.~\ref{fig:pareto_front_latency} shows the trade-offs between throughput and delay. Fig.~\ref{fig:pareto_front_throughput} shows the trade-offs between throughput and power consumption.}
    \label{fig:pareto_front}
\end{figure*}
\subsection{Baselines}
We compare the performance of our approach with the following baselines:
\begin{enumerate}
    \item We use the Orchestra~\cite{duquennoy2015orchestra} implementation in the Contiki-NG operating system~\cite{oikonomou2022contiki} as a baseline.
          Orchestra autonomously schedules the \acrshort{tsch} links with little overhead and high reliability.
    \item \highlight{We utilize the \acrfull{msf}, an established \acrfull{ietf} standard for \acrshort{tsch} scheduling, as a baseline. The \acrshort{msf} orchestrates the \acrshort{tsch} links within a unified shared cell, optimizing for minimal connectivity~\cite{chang20196tisch}.}
    \item \highlight{
              We also compare our approach with the QL-based TSCH scheduler~\cite{park2020multi} as a baseline.
          }
\end{enumerate}

\highlight{
    The configurations and notations of the protocols are summarized in Table~\ref{tab:protocol_parameters}.
    This table provides details such as the Slotframe Size (SF) utilized in the TSCH schedule, the optimized $\varphi$ combination for enhancing network performance, and whether retransmissions are enabled.
    The selection of SF for MSF and QL-TSCH is based on the required number of slots to accommodate network traffic.
    The default number of retransmissions is set to 3.
}

\begin{table}[ht!]
    \centering
    \caption{\highlight{Protocol parameters.}}
    \label{tab:protocol_parameters}
    \def\arraystretch{1}%
    \scriptsize
    \begin{NiceTabular}[c]{lcccl}[hvlines,cell-space-limits=1pt]
        \CodeBefore
        \rowcolor{lightgray}{1}
        \rowcolors{2}{gray!12}{}[respect-blocks]
        \Body
        \RowStyle[]{\bfseries}
        Protocol  & SF & $\varphi$       & ReTx   & Notation                     \\
        HRL-TSCH  & 17 & (0.5, 0.3, 0.2) & \xmark & HRL-TSCH($\varphi_1$)        \\
        HRL-TSCH  & 17 & (0.1, 0.5, 0.4) & \xmark & HRL-TSCH($\varphi_2$)        \\
        HRL-TSCH  & 17 & (0.2, 0.2, 0.6) & \xmark & HRL-TSCH($\varphi_3$)        \\
        HRL-TSCH  & 17 & (0.8, 0.1, 0.1) & \xmark & HRL-TSCH($\varphi_4$)        \\
        HRL-TSCH  & 17 & (0.1, 0.8, 0.1) & \xmark & HRL-TSCH($\varphi_5$)        \\
        HRL-TSCH  & 17 & (0.1, 0.1, 0.8) & \xmark & HRL-TSCH($\varphi_6$)        \\
        HRL-TSCH  & 17 & (0.5, 0.3, 0.2) & \cmark & HRL-TSCH($\varphi_1^{ReTx}$) \\
        HRL-TSCH  & 17 & (0.1, 0.5, 0.4) & \cmark & HRL-TSCH($\varphi_2^{ReTx}$) \\
        HRL-TSCH  & 17 & (0.2, 0.2, 0.6) & \cmark & HRL-TSCH($\varphi_3^{ReTx}$) \\
        HRL-TSCH  & 17 & (0.8, 0.1, 0.1) & \cmark & HRL-TSCH($\varphi_4^{ReTx}$) \\
        HRL-TSCH  & 17 & (0.1, 0.8, 0.1) & \cmark & HRL-TSCH($\varphi_5^{ReTx}$) \\
        HRL-TSCH  & 17 & (0.1, 0.1, 0.8) & \cmark & HRL-TSCH($\varphi_6^{ReTx}$) \\
        Orchestra & 11 & -               & \xmark & Orchestra-11                 \\
        Orchestra & 11 & -               & \cmark & Orchestra-11-ReTx            \\
        MSF       & 3  & -               & \xmark & MSF-3                        \\
        MSF       & 3  & -               & \cmark & MSF-3-ReTx                   \\
        MSF       & 5  & -               & \xmark & MSF-5                        \\
        MSF       & 5  & -               & \cmark & MSF-5-ReTx                   \\
        QL-TSCH-3 & 3  & -               & \xmark & QL-TSCH-3                    \\
        QL-TSCH-3 & 3  & -               & \cmark & QL-TSCH-3-ReTx               \\
        QL-TSCH-5 & 5  & -               & \xmark & QL-TSCH-5                    \\
        QL-TSCH-5 & 5  & -               & \cmark & QL-TSCH-5-ReTx               \\
    \end{NiceTabular}
\end{table}

\subsection{Pareto Front}
The Pareto front is constructed by systematically varying the user requirements, employing a finely tuned step size of 0.1 for precision.
We conducted 66 distinct simulations, each with a unique $\varphi$ combination.
Each simulation lasted 40 minutes, yielding over $16k$ packets. Refer to Fig.~\ref{fig:pareto_front} for a visualization of the power consumption, delay, and throughput trade-offs.
\highlight{
    To benchmark against the baselines, our strategy focuses on prioritizing $\varphi$ combinations that optimize trade-offs between power consumption and delay, throughput and delay, and power consumption and throughput.
    The $\varphi$ values that achieve the most favorable balance in these trade-offs are $\varphi_1=(0.5,0.3,0.2)$, $\varphi_2=(0.1,0.5,0.4)$, and $\varphi_3=(0.2,0.2,0.6)$, respectively.
    Moreover, we explore scenarios where one objective is emphasized over the others, represented by $\varphi_4=(0.8,0.1,0.1)$, $\varphi_5=(0.1,0.8,0.1)$, and $\varphi_6=(0.1,0.1,0.8)$.
}

\begin{figure}
    \centering
    \includegraphics[width=0.9\linewidth]{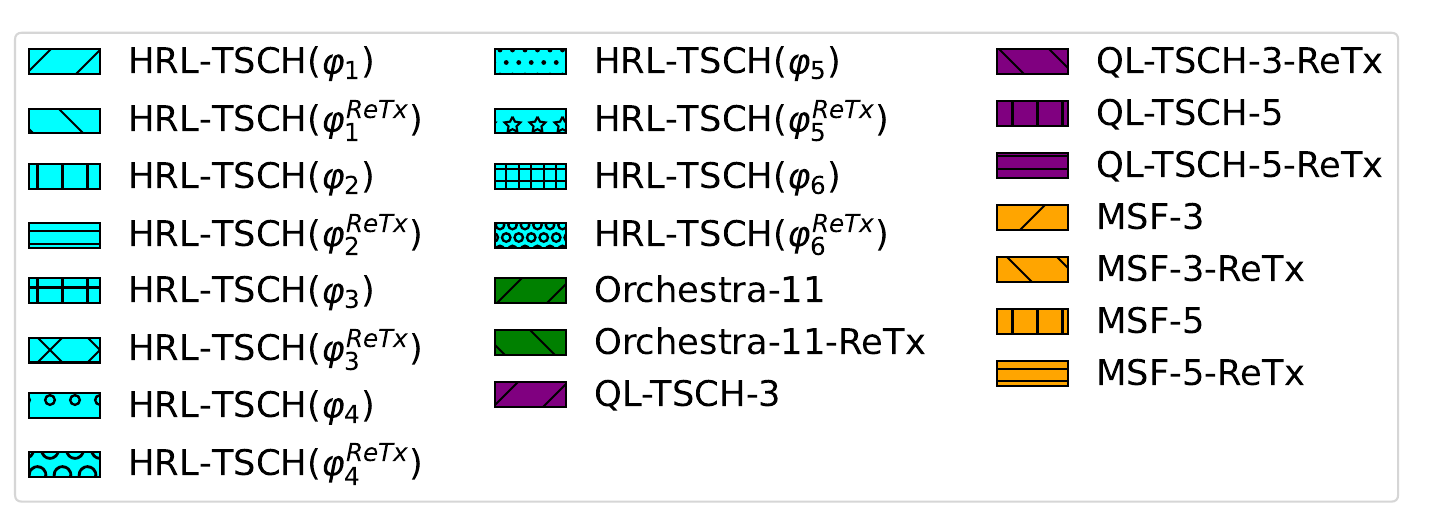}
    \caption{\highlight{Common legend for Fig.~\ref{fig:performance}, Fig.~\ref{fig:plr}, Fig.~\ref{fig:node_power}, Fig.~\ref{fig:node_throughput}, Fig.~\ref{fig:node_jitter} and Fig.~\ref{fig:node_packet_loss}.}}
    \label{fig:legend}
\end{figure}

\begin{figure*}
    \centering
    \subfloat[\label{fig:power_consumption}]{\includegraphics[width=0.32\textwidth]{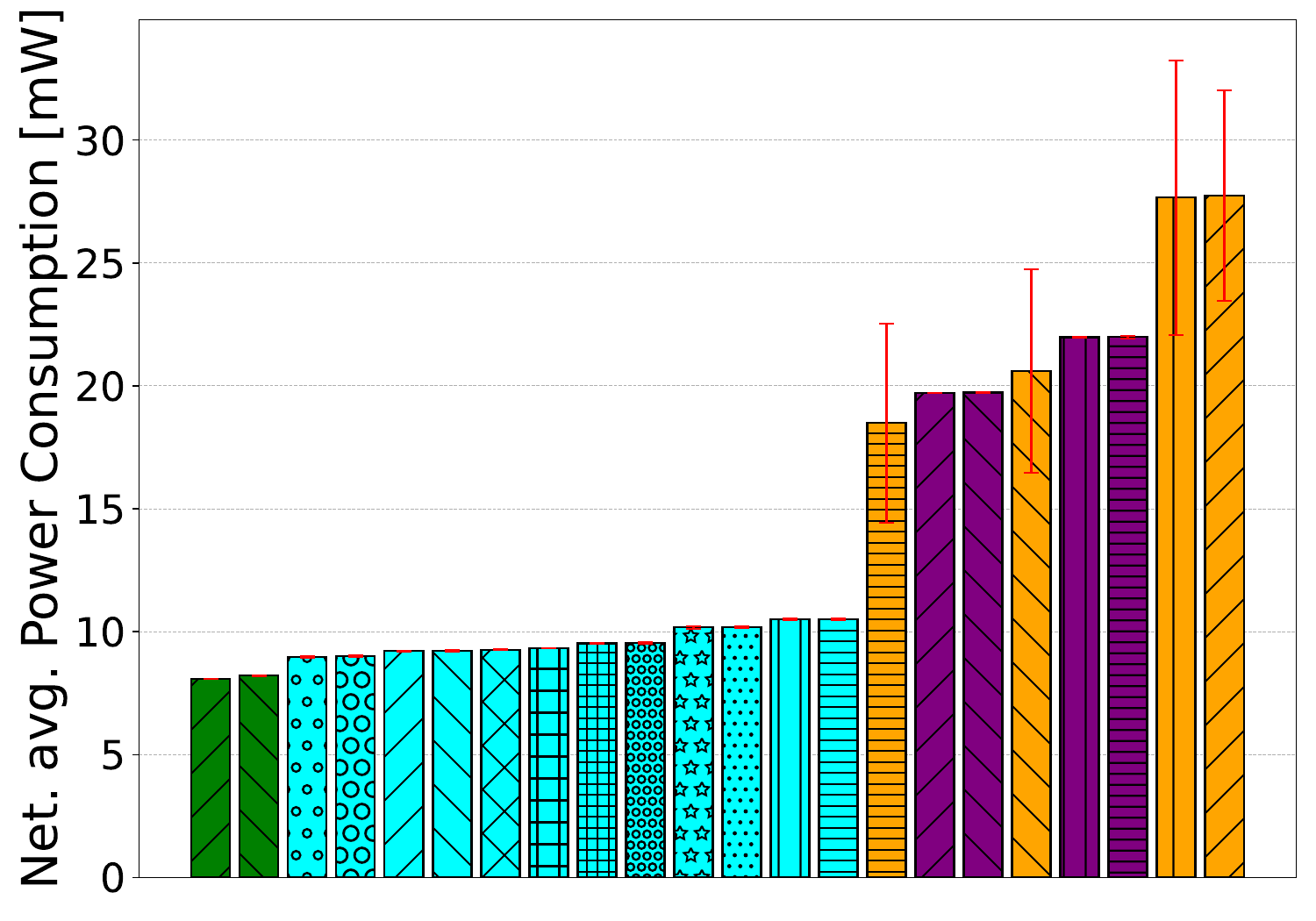}}\hfill
    \subfloat[\label{fig:latency}]{\includegraphics[width=0.32\textwidth]{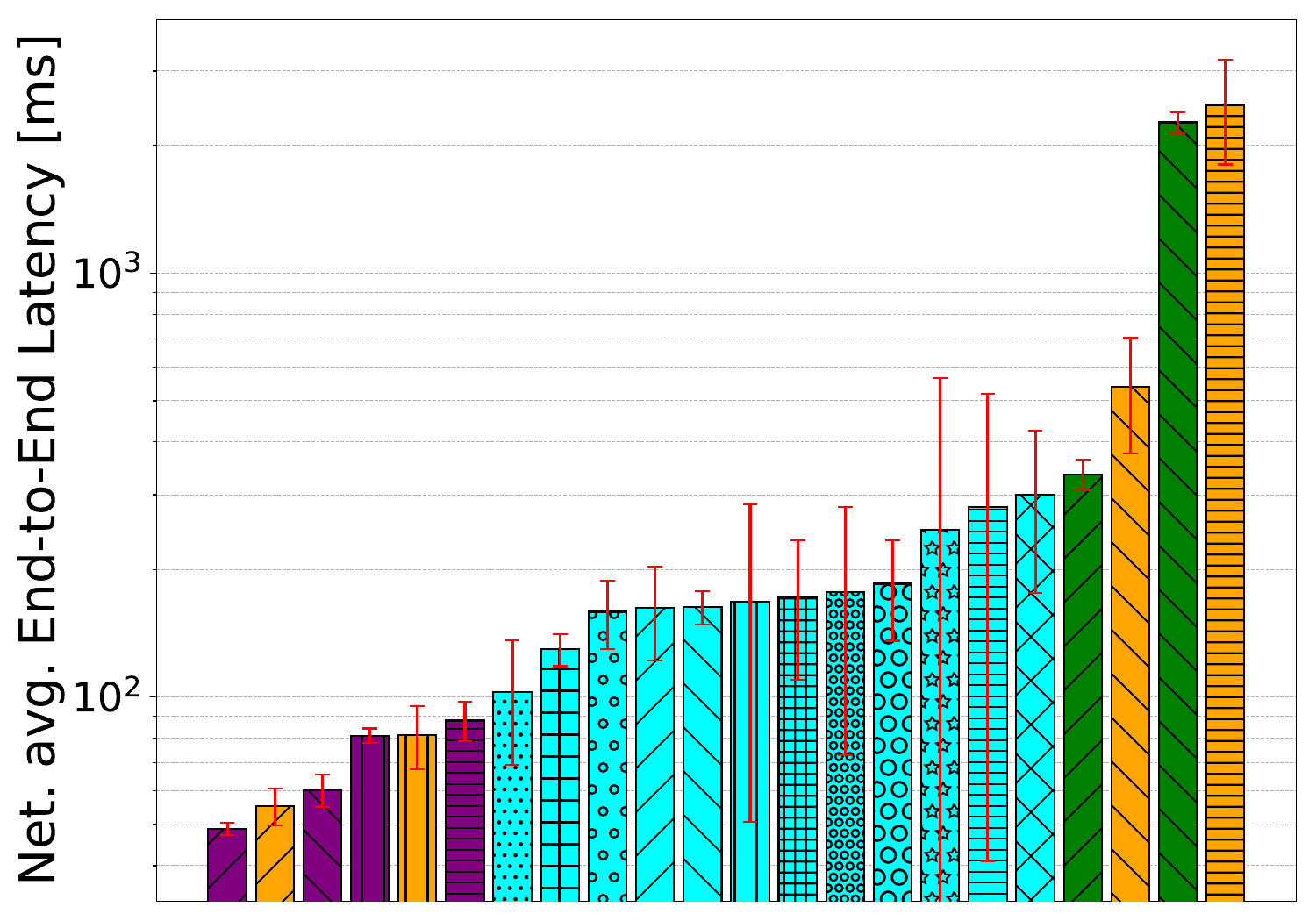}}\hfill
    \subfloat[\label{fig:throughput}]{\includegraphics[width=0.32\textwidth]{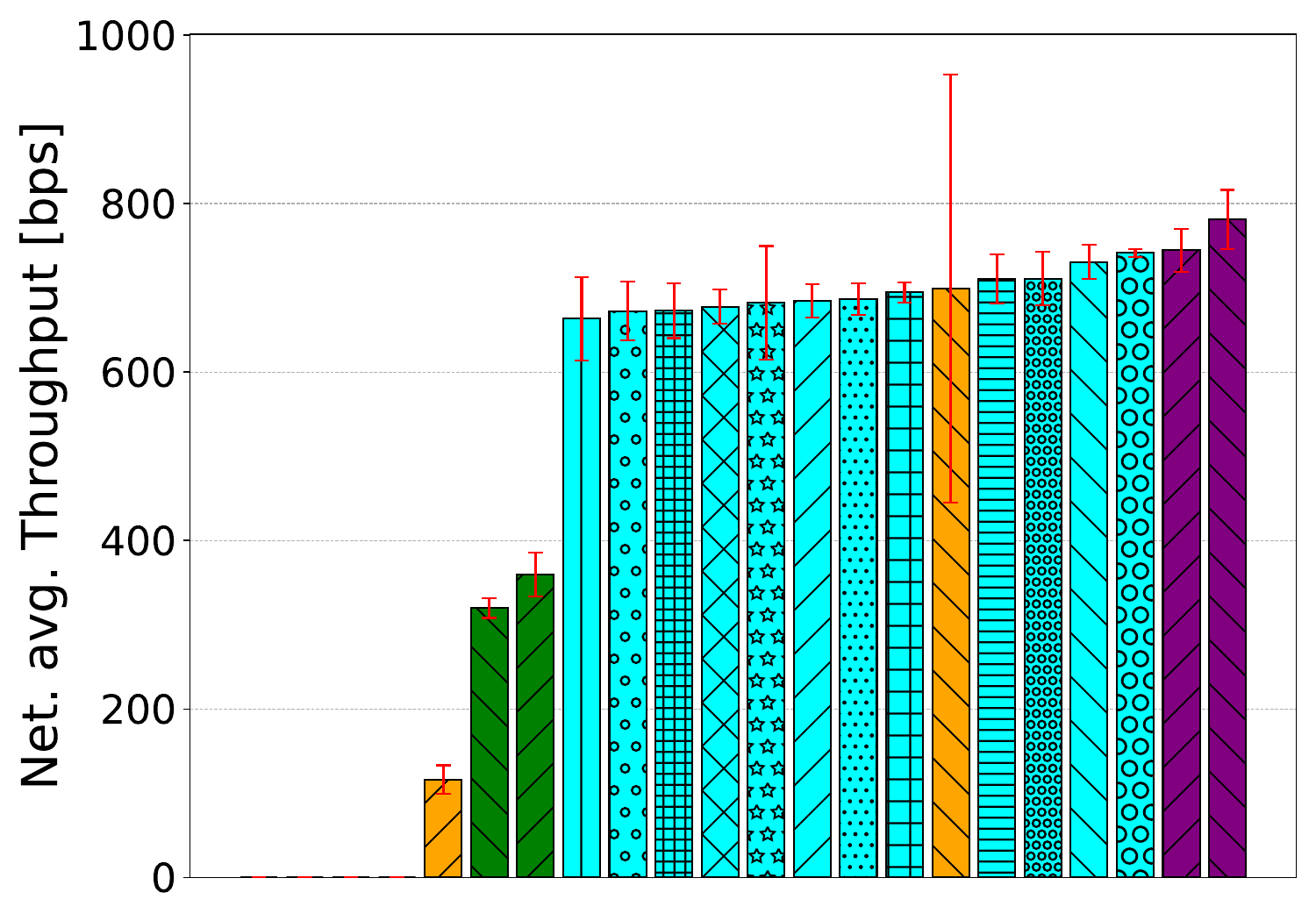}}
    \caption{\highlight{Performance comparison of \acrshort{hrltsch} and baselines for network power consumption, delay, and throughput (see Fig.~\ref{fig:legend} for the legend). (a) shows the average power consumption of the network. (b) shows the average delay of the network. (c) shows the average throughput of the network.}}
    \label{fig:performance}
\end{figure*}

\begin{figure}
    \centering
    \includegraphics[width=0.7\linewidth]{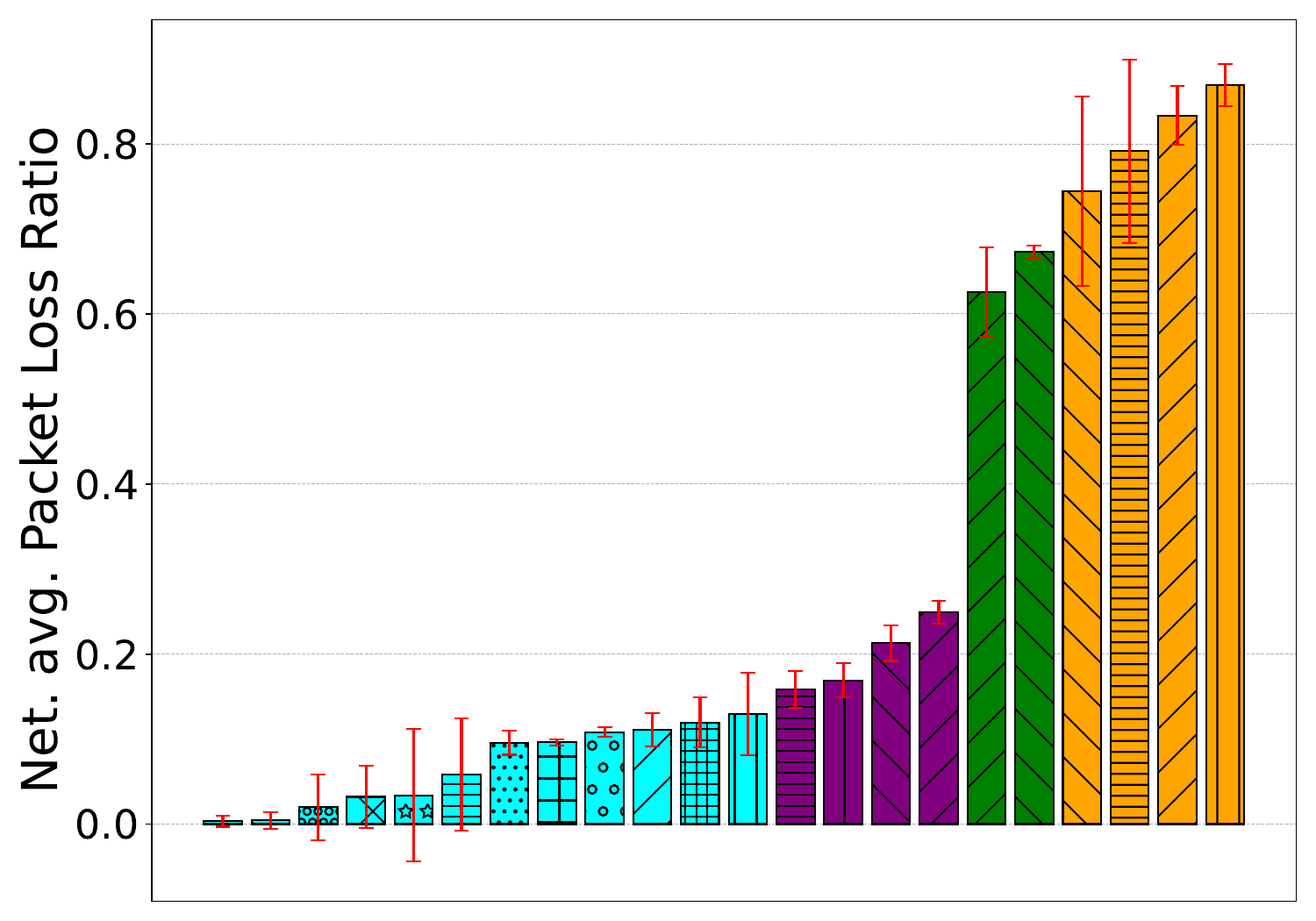}
    \caption{\highlight{\acrshort{plr} of \acrshort{hrltsch} and baselines (see Fig.~\ref{fig:legend} for the legend).}}
    \label{fig:plr}
\end{figure}

\begin{figure*}
    \centering
    \includegraphics[width=0.98\linewidth]{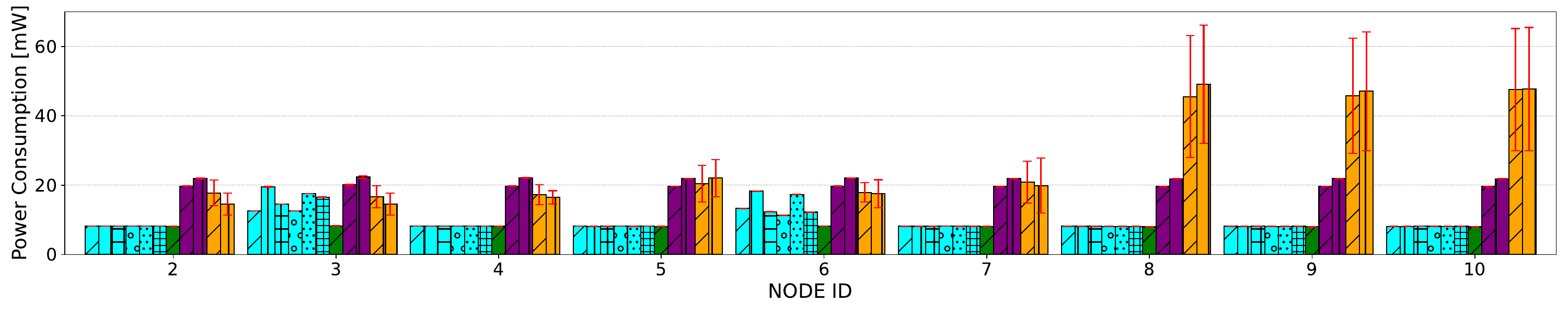}
    \caption{\highlight{Power consumption of \acrshort{hrltsch} and baselines for each node with retransmissions disabled (see Fig.~\ref{fig:legend} for the legend).}}
    \label{fig:node_power}
\end{figure*}

\begin{figure*}
    \centering
    \subfloat[\label{fig:latency_varphi1_per_node}]{\includegraphics[width=0.32\textwidth]{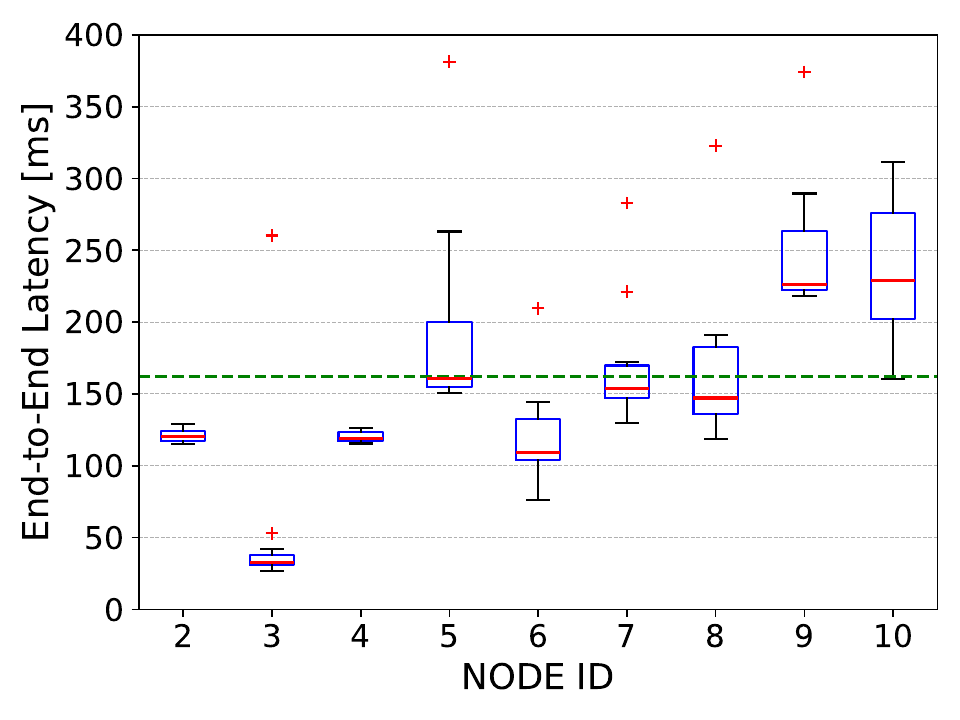}}\hfill
    \subfloat[\label{fig:latency_varphi2_per_node}]{\includegraphics[width=0.32\textwidth]{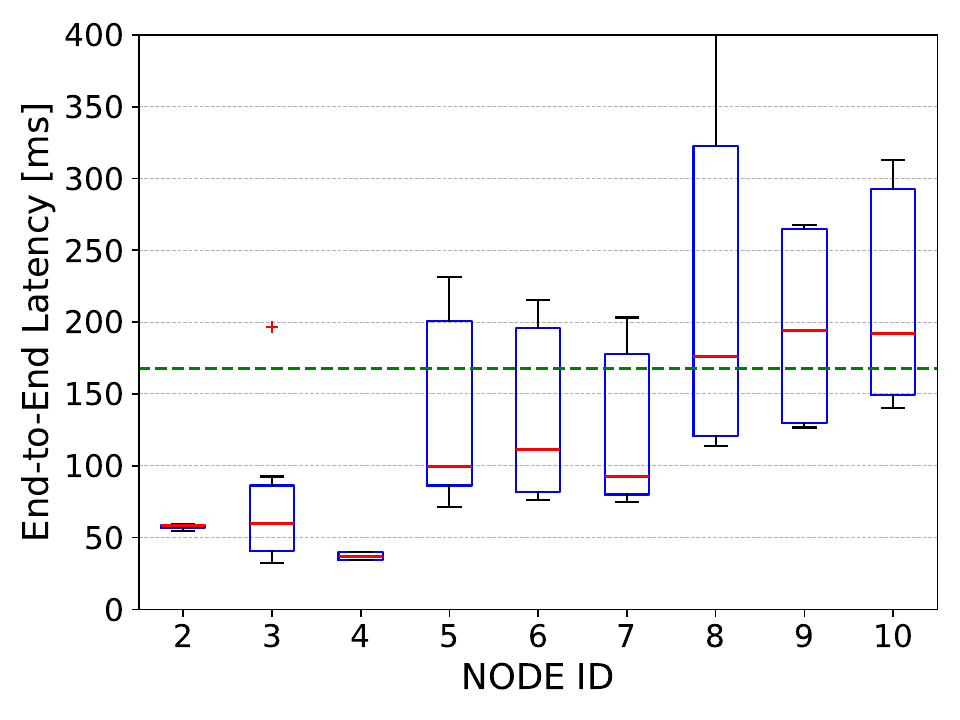}}\hfill
    \subfloat[\label{fig:latency_varphi3_per_node}]{\includegraphics[width=0.32\textwidth]{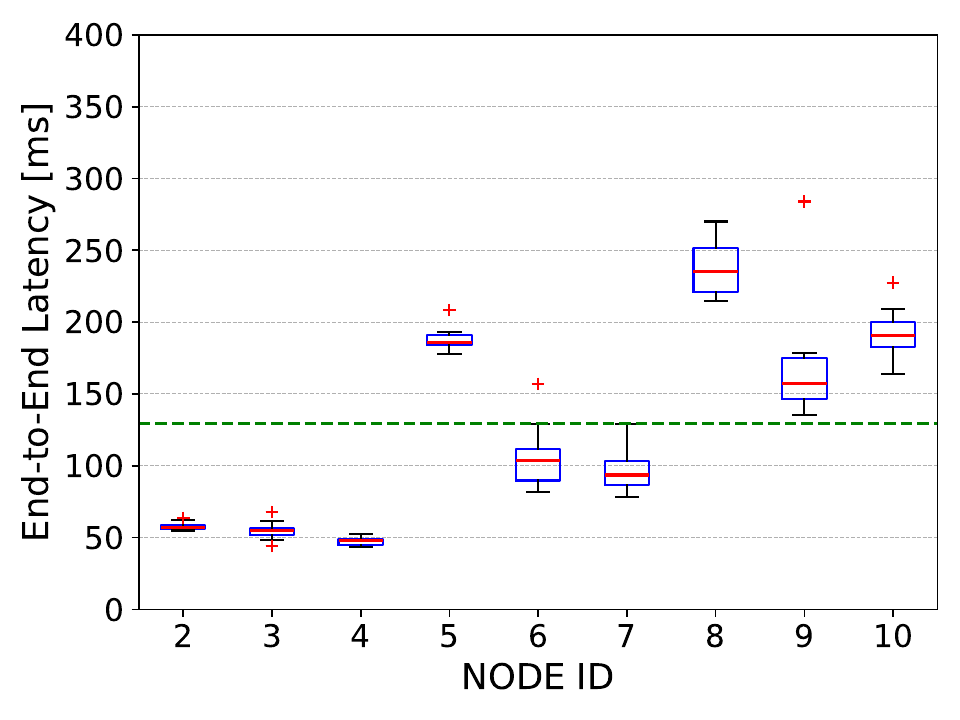}}\vfill
    \subfloat[\label{fig:latency_varphi4_per_node}]{\includegraphics[width=0.32\textwidth]{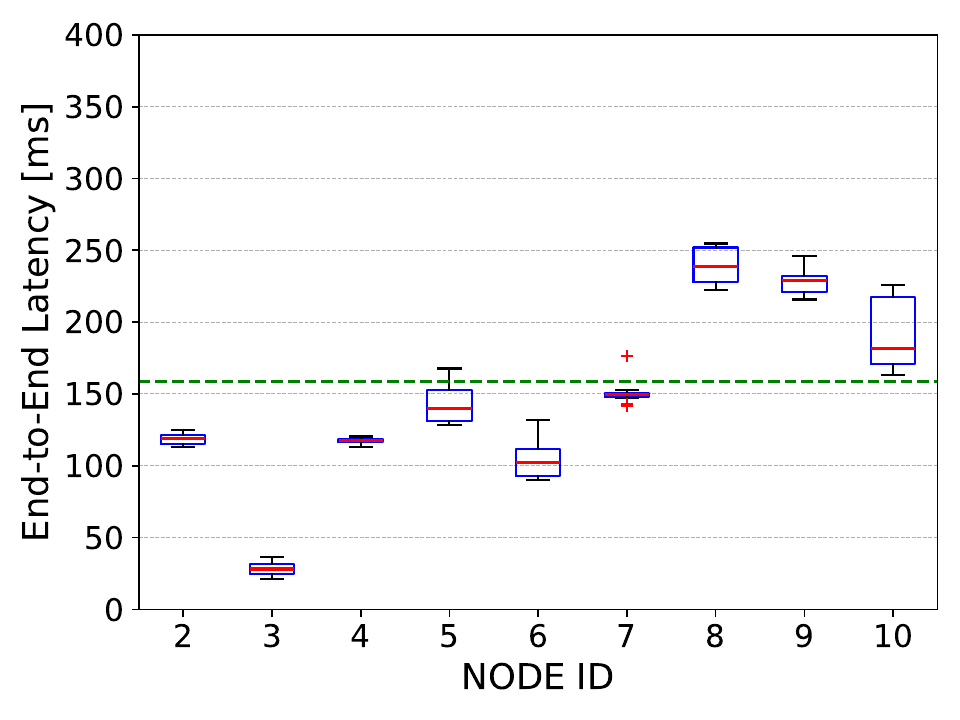}}\hfill
    \subfloat[\label{fig:latency_varphi5_per_node}]{\includegraphics[width=0.32\textwidth]{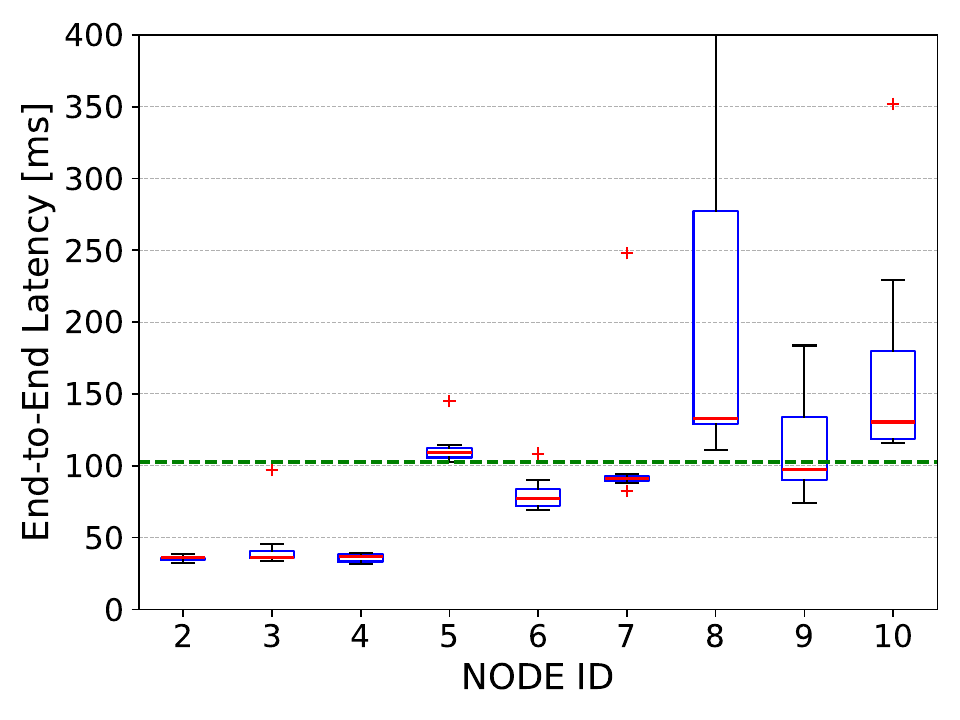}}\hfill
    \subfloat[\label{fig:latency_varphi6_per_node}]{\includegraphics[width=0.32\textwidth]{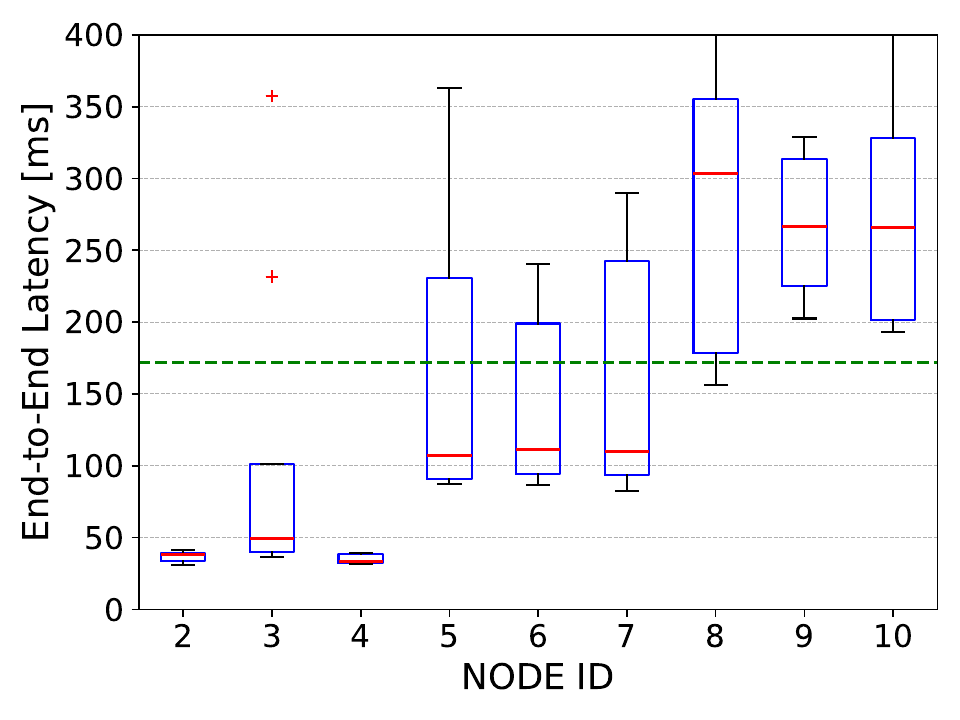}}
    \caption{\highlight{End-to-end latency of \acrshort{hrltsch} for each node with retransmission disabled. (a) $\varphi_1=(0.5,0.3,0.2)$. (b) $\varphi_2=(0.1,0.5,0.4)$. (c) $\varphi_3=(0.2,0.2,0.6)$. (d) $\varphi_4=(0.8,0.1,0.1)$. (e) $\varphi_5=(0.1,0.8,0.1)$. (f) $\varphi_6=(0.1,0.1,0.8)$.}}
    \label{fig:node_latency_per_node}
\end{figure*}

\subsection{Comparison with Baselines}

\subsubsection{Network power Consumption}

Fig. \ref{fig:power_consumption} illustrates the network's average power consumption.
\highlight{
    Among all protocols, \acrshort{msf} and QL-TSCH exhibit the highest power consumption, approximately two to three times higher than \acrshort{hrltsch}.
    \acrshort{msf} requires sensor nodes to remain in an active state throughout the network operation, while QL-TSCH's action peeking algorithm schedules receiving slots in all slots except the transmitting slot.
    Notably, \acrshort{hrltsch} achieves lower power consumption for the $\varphi_4$ combination compared to the other $\varphi$ combinations, as it prioritizes power consumption over delay and throughput.
}

\highlight{
    Given that \acrshort{hrltsch} allocates some attention to delay and throughput across all $\varphi$ combinations, it tends to consume more power than Orchestra, a correlation directly influenced by the requirements of delay and throughput.
    However, it's crucial to underscore that the substantial performance enhancements realized with \acrshort{hrltsch} justify its slightly higher power consumption.
}
Orchestra, conversely, optimizes consumption through receiver-based scheduling, employing fewer slots compared to \acrshort{hrltsch}.
\highlight{
    Despite this efficiency, Orchestra's approach significantly increases its \acrfull{plr} to approximately 65\% (almost five times higher than \acrshort{hrltsch}), contrasting with \acrshort{hrltsch}'s lower \acrshort{plr} of approximately 12\% for the worst $\varphi$ combination, as depicted in Fig. \ref{fig:plr}.
    Furthermore, it's noteworthy that \acrshort{hrltsch} operates in a contention-free manner, unlike Orchestra.
}

\subsubsection{Power Consumption per Node}

\highlight{
    Fig. \ref{fig:node_power} illustrates the power consumption per node for \acrshort{hrltsch} and the baseline protocols.
    Among the baselines, \acrshort{msf} and QL-TSCH demonstrate the highest power consumption per node.
    Specifically, QL-TSCH consumes approximately twice as much power per node compared to \acrshort{hrltsch}.
    Similarly, \acrshort{msf} exhibits roughly twice the power consumption per node compared to \acrshort{hrltsch}, except for nodes eight, nine, and ten, where \acrshort{msf} consumes approximately four times as much power.
    In contrast, HRL-TSCH demonstrates a more balanced power consumption per node across all nodes, with only some deviations observed at nodes three and six.
    These nodes are tasked with forwarding packets from deeper nodes within the network.
}

\subsubsection{End-to-End Latency}

Fig. \ref{fig:latency} illustrates the average delay experienced by the network, revealing distinct performance characteristics.
Notably, \highlight{
    protocols with the lowest slotframe size, such as MSF-3 and QL-TSCH-3, demonstrate the lowest delay among all protocols.
    This outcome is unsurprising, given that a lower slotframe size entails fewer slots to traverse, thereby reducing delay.
    However, this efficiency comes at the expense of higher power consumption, as indicated in Fig.~\ref{fig:power_consumption}.
    Additionally, it's important to highlight that while \acrshort{msf} and QL-TSCH achieve the lowest delay when a packet successfully traverses all hops in the network, this occurrence is infrequent, as shown in Fig. \ref{fig:plr}.
}

Specifically, \acrshort{hrltsch} outperforms Orchestra by achieving a consistently lower delay.
This superiority can be attributed to \acrshort{hrltsch}'s capacity to dynamically adapt the \acrshort{tsch} schedule in response to changing network conditions—an adaptability that Orchestra lacks.
This adaptability ensures that \acrshort{hrltsch} remains effective in minimizing delays across varying network states, contributing to its enhanced performance in comparison to Orchestra.

\subsubsection{\highlight{Analysis of Per-Node End-to-End Latency for HRL-TSCH}}

\highlight{
    Fig. \ref{fig:node_latency_per_node} illustrates the end-to-end delay per node for each $\varphi$ combination.
    All charts share the same scale to facilitate comparison.
    The dashed green line represents the average delay of the network.
    The $\varphi_5$ combination achieves the lowest delay for all nodes, as it prioritizes delay over power consumption and throughput.
    The $\varphi_3$ combination achieves the second-lowest delay for all nodes, as it prioritizes throughput over power consumption and delay, thus scheduling more slots for data transmission.
    The highest delay and deviation are observed for nodes eight, nine, and ten, located at the edge of the network, requiring more hops to reach the sink node.
}

\begin{figure*}
    \centering
    \includegraphics[width=0.98\linewidth]{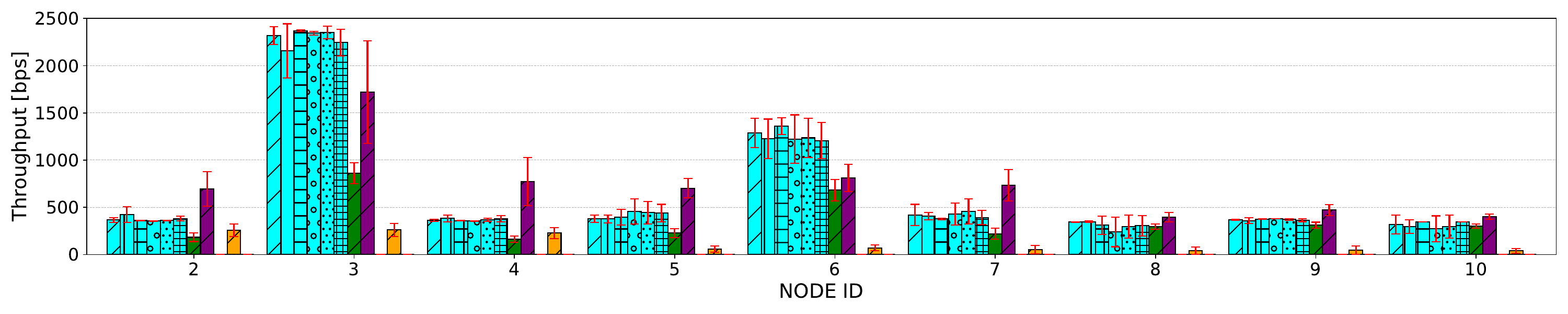}
    \caption{\highlight{Throughput of \acrshort{hrltsch} and baselines for each node with retransmissions disabled (see Fig.~\ref{fig:legend} for the legend).}}
    \label{fig:node_throughput}
\end{figure*}

\subsubsection{Throughput}

The average throughput of the network, depicted in Fig. \ref{fig:throughput}, highlights distinct performance advantages among the scheduling approaches.
Notably, \acrshort{hrltsch} emerges as the top performer, consistently achieving higher throughput compared to Orchestra, \highlight{\acrshort{msf}, and QL-TSCH.}

Orchestra lags behind in throughput performance (approximately twice lower than \acrshort{hrltsch}), primarily attributed to its receiver-based scheduling approach.
This method imposes limitations on the number of receiving slots within the \acrshort{tsch} schedule, constraining the overall throughput potential.
\highlight{Furthermore, \acrshort{msf}-3 and \acrshort{msf}-5 demonstrate poor throughput performance.
    The shared cell approach in \acrshort{msf} limits concurrent transmissions, resulting in lower throughput compared to \acrshort{hrltsch}.
    Conversely, QL-TSCH-3 and QL-TSCH-3-ReTx achieve the highest throughput among all \acrshort{msf} and QL-TSCH configurations.
    This is attributed to QL-TSCH's scheduling of receiving slots in all slots except the transmitting slot, facilitating increased concurrent transmissions.
    However, QL-TSCH-5 and QL-TSCH-5-ReTx exhibit the lowest throughput due to an increased slotframe size, which fails to accommodate the network traffic.
}
Consequently, \acrshort{hrltsch} stands out as the superior choice for achieving higher and more adaptable throughput in dynamic \acrshort{iiot} environments.

\subsubsection{\highlight{Analysis of Per-Node Throughput for HRL-TSCH}}

\highlight{
    Fig. \ref{fig:node_throughput} illustrates the throughput per node for each $\varphi$ combination as well as the baseline protocols.
    For all nodes except nodes three and six, all $\varphi$ combinations achieve similar throughput, slightly lower than that of QL-TSCH-3.
    Notably, HRL-TSCH consistently achieves the highest throughput for nodes three and six across all $\varphi$ combinations.
    This result aligns with expectations, as HRL-TSCH schedules more slots for data transmission compared to QL-TSCH-3, and nodes three and six typically handle more data traffic.
    Conversely, QL-TSCH-5 and \acrshort{msf}-5 consistently exhibit the lowest throughput for all nodes, indicating their inability to effectively manage network traffic demands.
}

\subsubsection{\highlight{Analysis of Per-Node Jitter}}

\begin{figure*}
    \centering
    \includegraphics[width=0.98\linewidth]{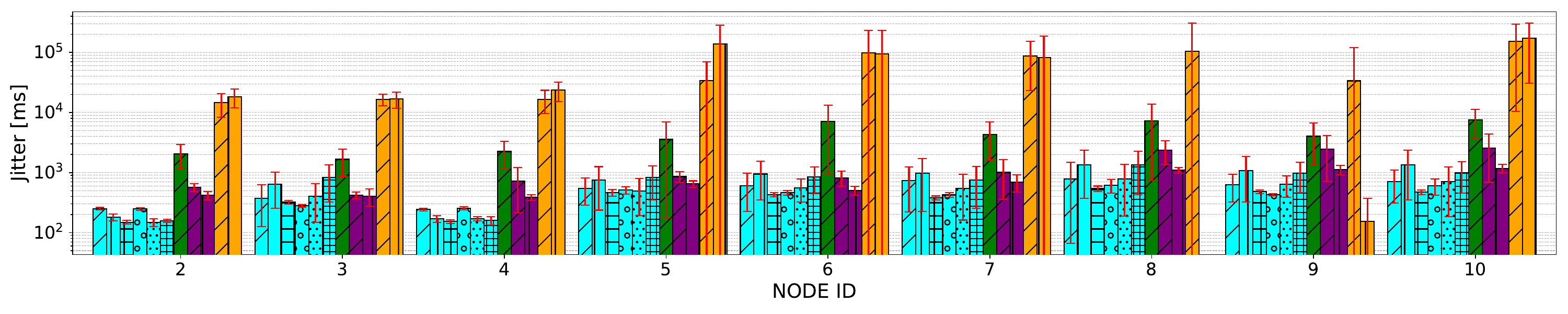}
    \caption{\highlight{Jitter of \acrshort{hrltsch} and baselines for each node with retransmissions disabled (see Fig.~\ref{fig:legend} for the legend).}}
    \label{fig:node_jitter}
\end{figure*}

\highlight{
    Fig. \ref{fig:node_jitter} presents the jitter analysis for each node.
    Jitter is computed as the time difference between the arrival of two consecutive packets.
    HRL-TSCH consistently outperforms other protocols across all $\varphi$ combinations in terms of jitter.
    Its meticulous scheduling approach ensures minimal delay in packet transmission and reception, resulting in consistently low jitter.
    Conversely, MSF-3 and MSF-5 exhibit the highest jitter among all protocols.
    This is primarily attributed to the shared cell approach in MSF, which restricts access to the shared cell, consequently leading to heightened jitter.
}

\subsubsection{\highlight{Analysis of Per-Node \acrshort{plr}}}

\begin{figure*}
    \centering
    \subfloat[\label{fig:plr_per_node}]{\includegraphics[width=0.98\textwidth]{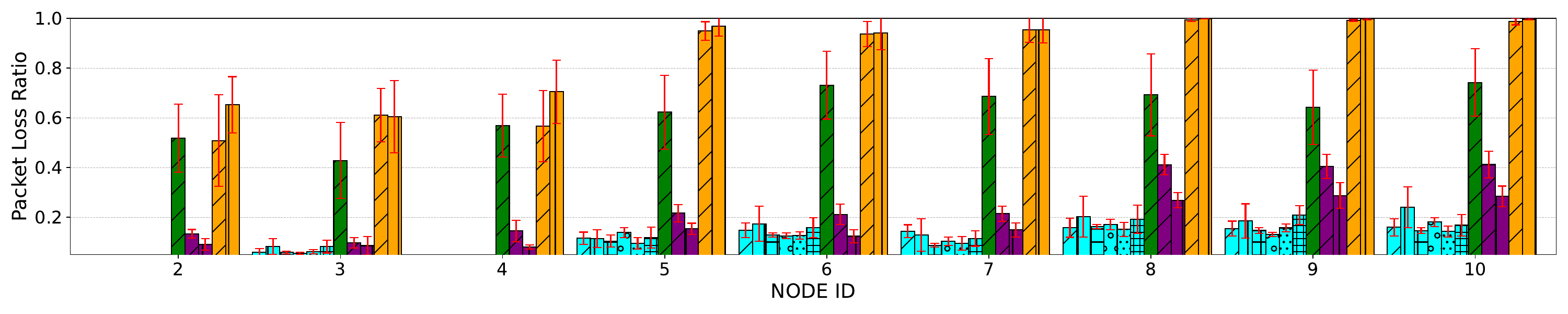}}\vfill
    \subfloat[\label{fig:plr_per_node_retx}]{\includegraphics[width=0.98\textwidth]{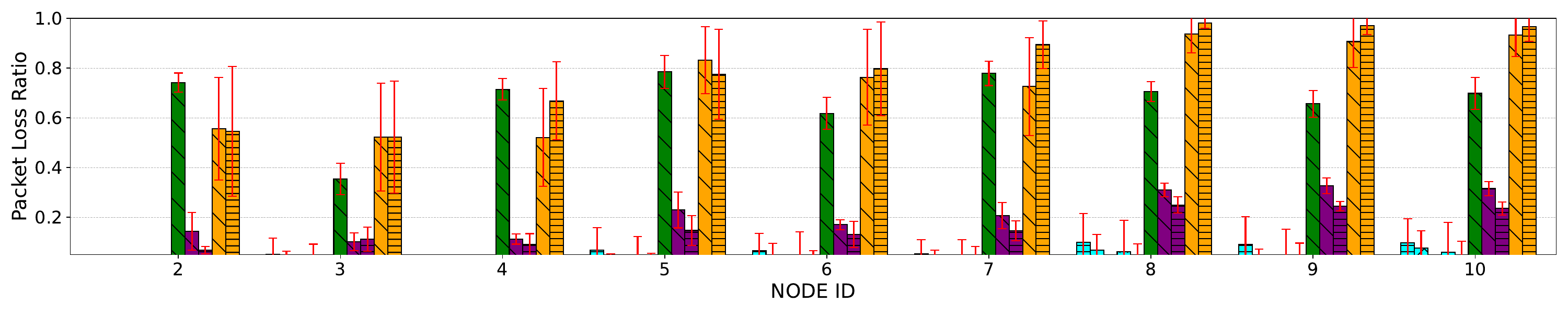}}\vfill
    \caption{\highlight{\acrshort{plr} of \acrshort{hrltsch} and baselines for each node (see Fig.~\ref{fig:legend} for the legend). (a) \acrshort{plr} without retransmissions. (b) \acrshort{plr} with retransmissions.}}
    \label{fig:node_packet_loss}
\end{figure*}

\highlight{
    The \acrshort{plr} analysis for each node is depicted in Fig. \ref{fig:node_packet_loss}. Fig. \ref{fig:plr_per_node} displays the \acrshort{plr} without retransmissions, while Fig. \ref{fig:plr_per_node_retx} showcases the \acrshort{plr} with retransmissions enabled.
    The \acrshort{plr} is calculated as the ratio of lost packets to the total number of transmitted packets.
    HRL-TSCH outperforms all baseline protocols in terms of \acrshort{plr} across all nodes.
    Particularly noteworthy is its superior performance at edge nodes, where the \acrshort{plr} tends to be higher due to the increased number of hops.
    Notably, the \acrshort{plr} is significantly higher for Orchestra, MSF, and QL-TSCH compared to HRL-TSCH, underlining the latter's effectiveness in minimizing packet loss.
    Furthermore, HRL-TSCH achieves the lowest \acrshort{plr} even without retransmissions enabled, further underscoring its robustness in maintaining packet integrity.
}

\begin{figure*}
    \centering
    \subfloat[\label{fig:power_vs_latency}]{\includegraphics[width=0.32\textwidth]{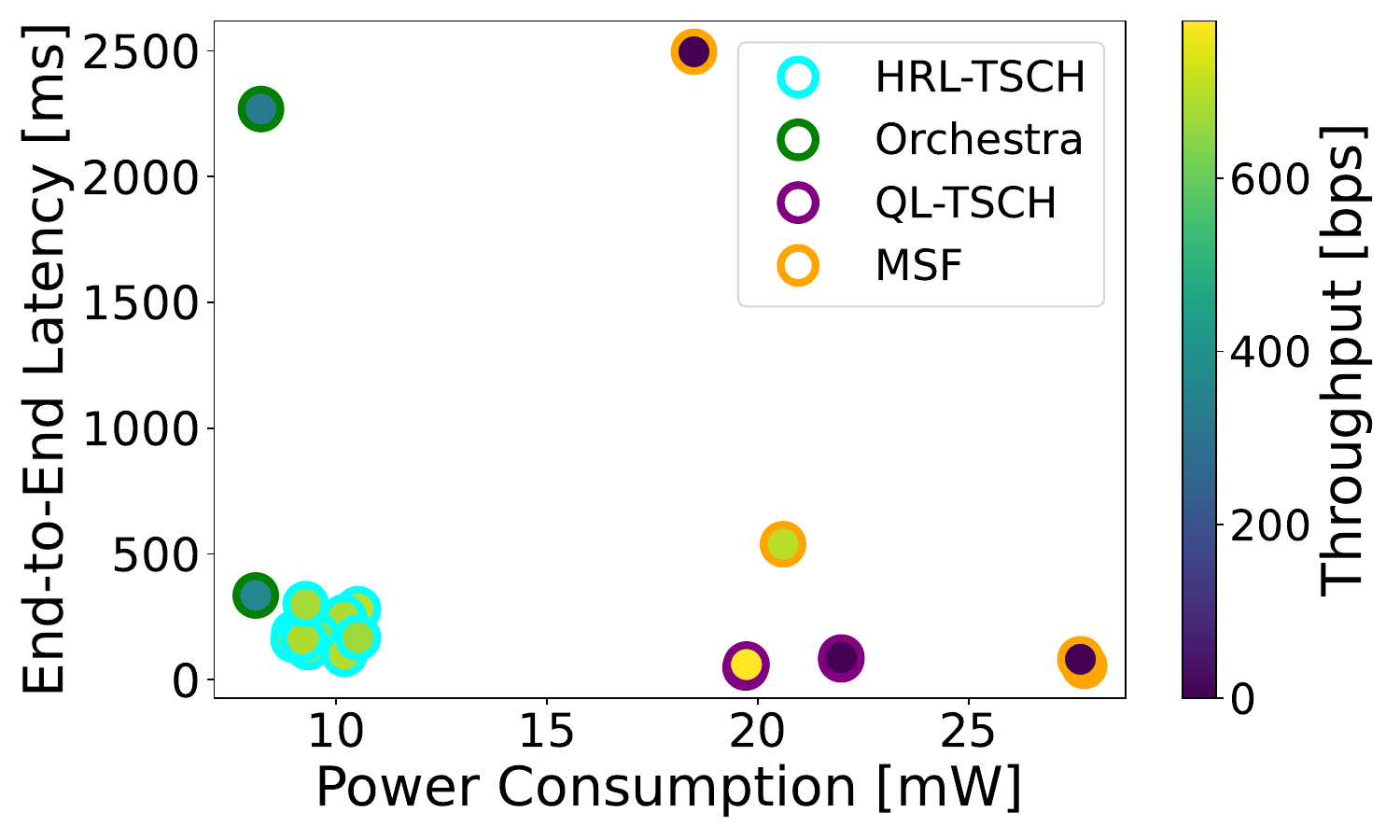}}\hfill
    \subfloat[\label{fig:power_vs_throughput}]{\includegraphics[width=0.32\textwidth]{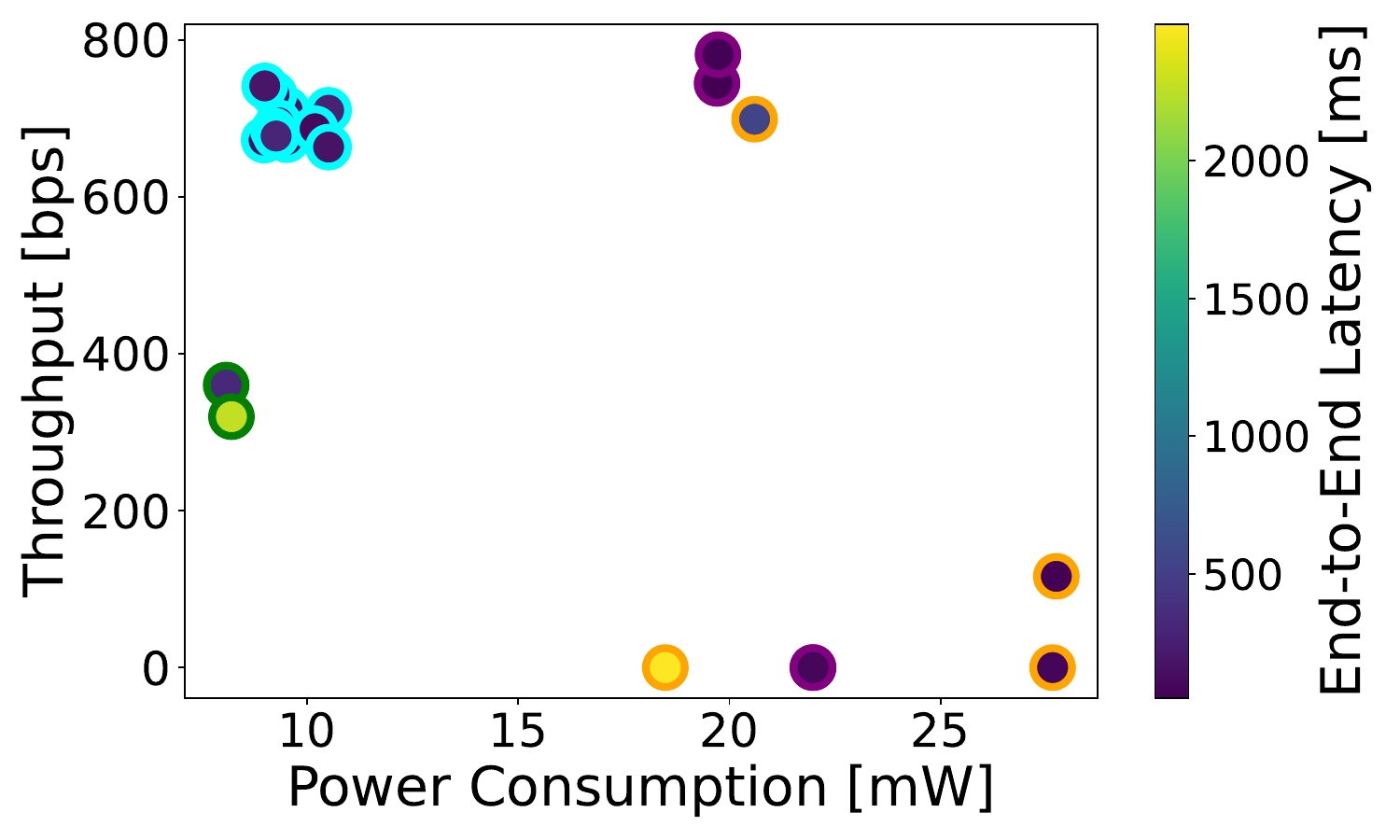}}\hfill
    \subfloat[\label{fig:latency_vs_throughput}]{\includegraphics[width=0.32\textwidth]{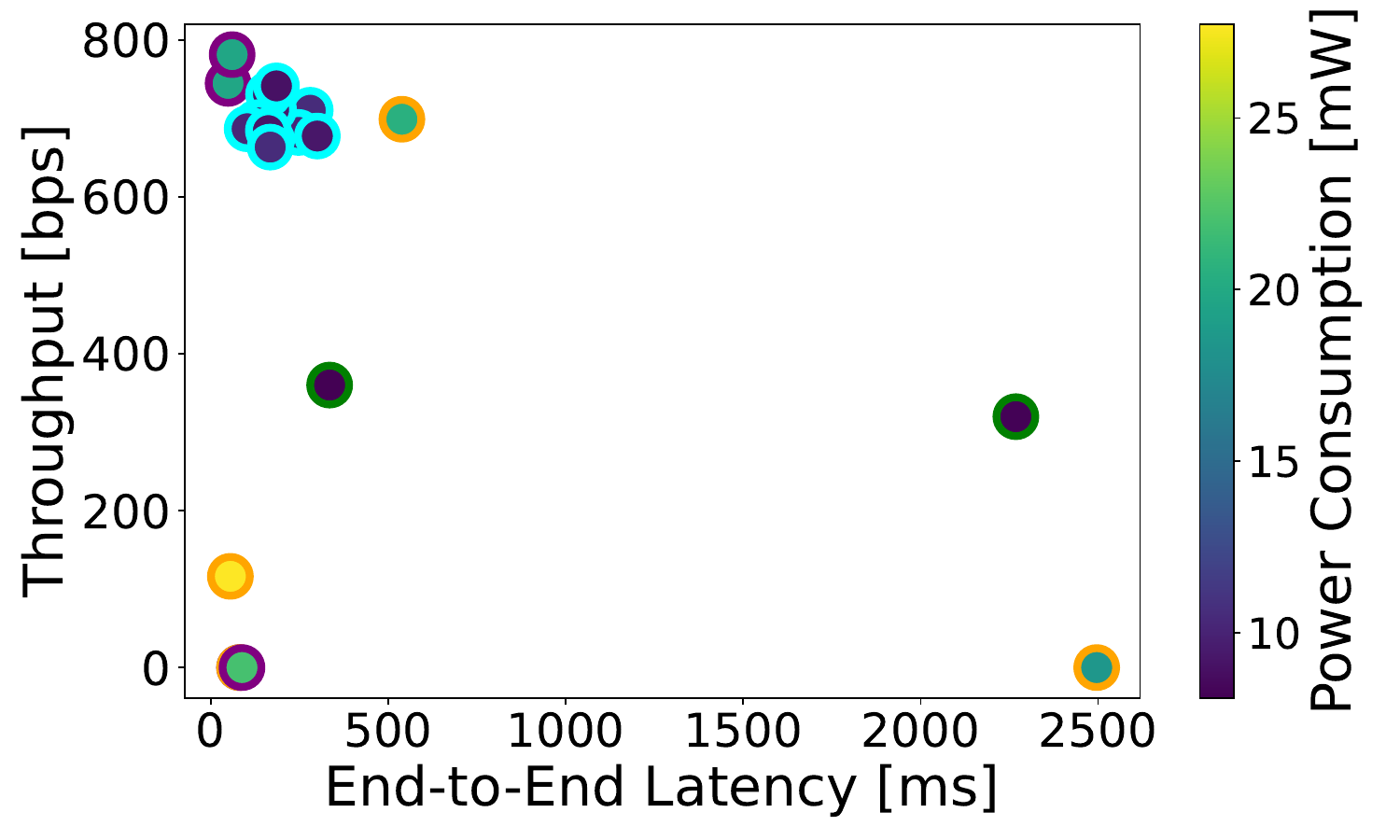}}
    \caption{\highlight{Heatmap that shows the trade-offs between power consumption, delay, and throughput. (a) shows the trade-offs between power consumption and delay. (b) shows the trade-offs between power consumption and throughput. (c) shows the trade-offs between throughput and delay.}}
    \label{fig:heatmap}
\end{figure*}

\begin{table*}[ht!]
    \centering
    \caption{\highlight{Ranking of protocols for a balanced, power-efficient, delay-sensitive, throughput-sensitive, and reliability-sensitive network.}}
    \label{tab:protocol_ranking}
    \def\arraystretch{1}%
    \fontsize{5.7}{7}\selectfont
    \begin{NiceTabular}[c]{clclclclcl}[hvlines,cell-space-limits=1pt]
        \CodeBefore
        \rowcolor{lightgray}{1,2}
        \rowcolors{2}{gray!12}{}[respect-blocks]
        \Body
        \RowStyle[]{\bfseries}
        \Block{2-1}{Balanced                                                                                                                                                                             \\(points)} & \Block{2-1}{Protocol} & \Block{2-1}{Power\\(points)} & \Block{2-1}{Protocol} & \Block{2-1}{Delay\\(points)} & \Block{2-1}{Protocol} & \Block{2-1}{Throughput\\(points)} & \Block{2-1}{Protocol} & \Block{2-1}{Reliability\\(points)} &\Block{2-1}{Protocol}\\
              &                              &       &                              &       &                              &       &                              &       &                              \\
        96.17 & HRL-TSCH($\varphi_4^{ReTx}$) & 95.68 & HRL-TSCH($\varphi_4^{ReTx}$) & 95.51 & HRL-TSCH($\varphi_1^{ReTx}$) & 98.47 & HRL-TSCH($\varphi_4^{ReTx}$) & 95.4  & HRL-TSCH($\varphi_4^{ReTx}$) \\
        95.76 & HRL-TSCH($\varphi_1^{ReTx}$) & 94.84 & HRL-TSCH($\varphi_1^{ReTx}$) & 95.13 & HRL-TSCH($\varphi_4^{ReTx}$) & 98.26 & HRL-TSCH($\varphi_1^{ReTx}$) & 94.42 & HRL-TSCH($\varphi_1^{ReTx}$) \\
        94.14 & HRL-TSCH($\varphi_6^{ReTx}$) & 93.75 & HRL-TSCH($\varphi_4$)        & 95.12 & HRL-TSCH($\varphi_5$)        & 96.54 & HRL-TSCH($\varphi_6^{ReTx}$) & 92.26 & HRL-TSCH($\varphi_6^{ReTx}$) \\
        92.15 & HRL-TSCH($\varphi_3$)        & 93.25 & HRL-TSCH($\varphi_6^{ReTx}$) & 94.88 & HRL-TSCH($\varphi_3$)        & 94.74 & HRL-TSCH($\varphi_3^{ReTx}$) & 91.61 & QL-TSCH-3-ReTx               \\
        91.78 & HRL-TSCH($\varphi_3^{ReTx}$) & 93.1  & HRL-TSCH($\varphi_3^{ReTx}$) & 94.51 & HRL-TSCH($\varphi_6^{ReTx}$) & 94.4  & HRL-TSCH($\varphi_5^{ReTx}$) & 90.84 & HRL-TSCH($\varphi_2^{ReTx}$) \\
        91.26 & HRL-TSCH($\varphi_5^{ReTx}$) & 93.08 & HRL-TSCH($\varphi_3$)        & 93.8  & HRL-TSCH($\varphi_4$)        & 92.51 & HRL-TSCH($\varphi_2^{ReTx}$) & 90.19 & HRL-TSCH($\varphi_3$)        \\
        91.23 & HRL-TSCH($\varphi_4$)        & 93.07 & HRL-TSCH($\varphi_1$)        & 93.7  & HRL-TSCH($\varphi_1$)        & 90.43 & HRL-TSCH($\varphi_3$)        & 89.18 & HRL-TSCH($\varphi_5$)        \\
        91.21 & HRL-TSCH($\varphi_1$)        & 91.65 & HRL-TSCH($\varphi_6$)        & 93.02 & HRL-TSCH($\varphi_6$)        & 90.03 & HRL-TSCH($\varphi_5$)        & 89.05 & HRL-TSCH($\varphi_1$)        \\
        91.08 & HRL-TSCH($\varphi_5$)        & 90.14 & HRL-TSCH($\varphi_5^{ReTx}$) & 92.41 & HRL-TSCH($\varphi_2$)        & 89.25 & HRL-TSCH($\varphi_4$)        & 88.89 & HRL-TSCH($\varphi_5^{ReTx}$) \\
        90.71 & HRL-TSCH($\varphi_2^{ReTx}$) & 90.01 & HRL-TSCH($\varphi_5$)        & 91.63 & HRL-TSCH($\varphi_5^{ReTx}$) & 89.03 & HRL-TSCH($\varphi_1$)        & 88.74 & HRL-TSCH($\varphi_3^{ReTx}$) \\
        90.09 & HRL-TSCH($\varphi_6$)        & 88.88 & HRL-TSCH($\varphi_2^{ReTx}$) & 91.33 & QL-TSCH-3-ReTx               & 87.99 & HRL-TSCH($\varphi_6$)        & 88.12 & HRL-TSCH($\varphi_4$)        \\
        88.29 & HRL-TSCH($\varphi_2$)        & 87.91 & HRL-TSCH($\varphi_2$)        & 90.78 & QL-TSCH-3                    & 86.58 & HRL-TSCH($\varphi_2$)        & 87.96 & QL-TSCH-3                    \\
        79.03 & QL-TSCH-3-ReTx               & 86.25 & Orchestra-11                 & 90.61 & HRL-TSCH($\varphi_2^{ReTx}$) & 77.11 & QL-TSCH-3-ReTx               & 87.71 & HRL-TSCH($\varphi_6$)        \\
        76.95 & QL-TSCH-3                    & 76.85 & Orchestra-11-ReTx            & 90.55 & HRL-TSCH($\varphi_3^{ReTx}$) & 73.74 & QL-TSCH-3                    & 86.26 & HRL-TSCH($\varphi_2$)        \\
        65.62 & Orchestra-11                 & 56.07 & QL-TSCH-3-ReTx               & 80.1  & QL-TSCH-5                    & 70.28 & QL-TSCH-5-ReTx               & 75.69 & MSF-3-ReTx                   \\
        55.05 & MSF-3-ReTx                   & 55.3  & QL-TSCH-3                    & 80.03 & QL-TSCH-5-ReTx               & 69.41 & QL-TSCH-5                    & 53.9  & Orchestra-11                 \\
        52.46 & QL-TSCH-5-ReTx               & 43.83 & MSF-3-ReTx                   & 79.24 & Orchestra-11                 & 43.1  & Orchestra-11                 & 41.79 & Orchestra-11-ReTx            \\
        52.22 & QL-TSCH-5                    & 38.57 & QL-TSCH-5-ReTx               & 71.71 & MSF-3                        & 30.83 & Orchestra-11-ReTx            & 20.99 & QL-TSCH-5-ReTx               \\
        43.07 & Orchestra-11-ReTx            & 38.49 & QL-TSCH-5                    & 70.01 & MSF-3-ReTx                   & 30.66 & MSF-3-ReTx                   & 20.89 & QL-TSCH-5                    \\
        29.68 & MSF-3                        & 33.86 & MSF-5-ReTx                   & 69.12 & MSF-5                        & 14.33 & MSF-3                        & 20.8  & MSF-3                        \\
        24.78 & MSF-5                        & 11.87 & MSF-3                        & 22.81 & Orchestra-11-ReTx            & 10.98 & MSF-5-ReTx                   & 9.91  & MSF-5                        \\
        14.01 & MSF-5-ReTx                   & 10.18 & MSF-5                        & 5.61  & MSF-5-ReTx                   & 9.91  & MSF-5                        & 5.61  & MSF-5-ReTx                   \\
    \end{NiceTabular}
\end{table*}

\subsubsection{\highlight{Protocol Ranking}}

In Fig. \ref{fig:heatmap}, a comprehensive representation of trade-offs between power consumption, delay, and throughput is depicted.
The proximity of a protocol to the minimum values of power consumption and delay, while maximizing throughput, serves as a crucial indicator of its overall performance in balancing these trade-offs.

\highlight{
    In this context, Fig.~\ref{fig:power_vs_latency} illustrates the trade-offs between power consumption and delay.
    Protocols positioned closer to the origin of the power consumption and delay axes are deemed optimal.
    Additionally, protocols depicted with a lighter shade of vivid color indicate higher throughput.
    It's evident that HRL-TSCH, across all $\varphi$ combinations, is positioned closer to the origin compared to the baseline protocols.
    Furthermore, they consistently exhibit higher throughput compared to the baselines.
}

\highlight{
    Fig.~\ref{fig:power_vs_throughput} visualizes the trade-offs between power consumption and throughput.
    In this representation, protocols closer to the y-axis signify lower power consumption, while those positioned further from the x-axis indicate higher throughput.
    Notably, HRL-TSCH configurations (clustered in the top-left quadrant) are situated closer to the y-axis and further from the x-axis compared to the baseline protocols.
    This positioning underscores that HRL-TSCH achieves superior throughput with lower power consumption relative to the baselines.
    Furthermore, across all $\varphi$ combinations, HRL-TSCH demonstrates lower delay, consolidating its performance advantages.
}

\highlight{
    Fig.~\ref{fig:latency_vs_throughput} illustrates the trade-offs between delay and throughput.
    In this depiction, protocols positioned closer to the y-axis signify lower delay, while those situated further from the x-axis indicate higher throughput.
    Notably, all HRL-TSCH $\varphi$ combinations are clustered in the top-left corner of the graph, indicating that HRL-TSCH achieves superior throughput with lower delay compared to the baseline protocols.
    Furthermore, across all $\varphi$ combinations, HRL-TSCH exhibits lower power consumption, further solidifying its performance advantages.
}

\highlight{
    To comprehensively evaluate the performance of each protocol, we adopt a multi-dimensional ranking approach, considering factors such as balanced performance, power efficiency, sensitivity to delay, throughput, and reliability within the network.
    Firstly, we assign scores to each protocol based on their performance across these metrics.
    We utilize linear interpolation to determine these scores using the following formula:
    \begin{equation}
        s_i = s_{\text{min}} + \frac{(s_{\text{max}} - s_{\text{min}}) \times (x_i - x_{\text{min}})}{(x_{\text{max}} - x_{\text{min}})}
    \end{equation}
    Here, $s_i$ represents the score of the protocol, with $s_{\text{min}}$ and $s_{\text{max}}$ denoting the minimum and maximum scores, respectively.
    $x_i$ denotes the value of the metric, while $x_{\text{min}}$ and $x_{\text{max}}$ represent the minimum and maximum values of the metric.
    The minimum and maximum scores are set to 0 and 100, or vice versa, depending on the nature of the metric.
    Subsequently, to calculate the final score for each protocol, we weigh the scores of each metric based on their relative importance.
    The final score ($S$) is computed as follows:
    \begin{equation}
        \begin{aligned}
        S = &\sum_{i=1}^{n} w_i \times s_i \\
        \text{where} \quad &\sum_{i=1}^{n} w_i = 1
        \end{aligned}
    \end{equation}
    Here, $w_i \in W$ represents the weight of the $i^{th}$ metric, while $n$ denotes the number of metrics.
    The weight vector $W$ comprises weights assigned to power consumption, delay, throughput, and reliability ($w_1$, $w_2$, $w_3$, and $w_4$, respectively).
    We consider the following weight configurations for each metric: balanced performance ($w_b=(0.25,0.25,0.25,0.25)$), power efficiency ($w_p=(0.7,0.1,0.1,0.1)$), delay sensitivity ($w_d=(0.1,0.7,0.1,0.1)$), throughput sensitivity ($w_t=(0.1,0.1,0.7,0.1)$), and reliability sensitivity ($w_r=(0.1,0.1,0.1,0.7)$).
}

\highlight{
    Table \ref{tab:protocol_ranking} displays the rankings of protocols for networks emphasizing balanced performance, power efficiency, delay sensitivity, throughput sensitivity, and reliability.
    These rankings are determined based on the final score assigned to each protocol, calculated using weighted scores across multiple performance metrics.
    In this ranking system, protocols with higher final scores are positioned at the top, signifying superior overall performance.
    Remarkably, the ranking remains consistent across various weight configurations, underscoring the robustness of HRL-TSCH across diverse network requirements.
    Notably, HRL-TSCH consistently secures the top position across all weighted scenarios, highlighting its versatility and efficacy across different network priorities.
    Whether the focus is on achieving a well-balanced performance or optimizing for specific criteria such as power efficiency, delay sensitivity, throughput sensitivity, or reliability, HRL-TSCH emerges as the optimal choice.
    This consistent dominance underscores the adaptability and reliability of HRL-TSCH in meeting the demands of modern networking applications, making it the preferred solution for a wide range of network configurations and objectives.
}

\section{Conclusion and Future Work}\label{sec:conclusion}
In this paper, we introduced \acrshort{hrltsch}, a novel hierarchical reinforcement learning framework tailored for optimizing the \acrshort{tsch} schedule within an \acrshort{iot} network, specifically designed to meet diverse application requirements.
Our approach leverages a higher-level policy for optimal link selection and a lower-level policy for optimal cell selection, both trained using the \acrshort{dqn} algorithm.
The inherent adaptability of our methodology allows it to effectively address a comprehensive range of application requirements, encompassing power consumption, delay, and throughput.

Our experiments, conducted in the Cooja network simulator, surely demonstrated that \acrshort{hrltsch} outperforms existing state-of-the-art approaches, affirming its crucial role in balancing trade-offs between power consumption, delay, and throughput.

Looking ahead, our future work will extend the optimization scope by incorporating slotframe size into the \acrshort{hrltsch} framework, recognizing the essential role of \acrshort{hrl} in further improving the performance of \acrshort{tsch} schedules.
Additionally, we plan to broaden the applicability of \acrshort{hrltsch} by extending support for contention-based scheduling, solidifying \acrshort{hrl}'s key role in enhancing the versatility and adaptability of \acrshort{iot} networks across diverse scenarios.
\highlight{
    While our study primarily focuses on optimizing the TSCH schedule to meet specific user requirements rather than explicitly addressing single-point-of-failure issues, future research could explore strategies for enhancing system robustness, such as incorporating redundancy or fault-tolerant mechanisms.
}
The nature of \acrshort{hrl} in solving the \acrshort{tsch} scheduling problem positions \acrshort{hrltsch} as a pivotal advancement in optimizing \acrshort{tsch} schedules for dynamic \acrshort{iot} environments.

\bibliographystyle{IEEEtran}
\bibliography{aux_files/IEEEabrv,references}

\begin{thebibliography}{10}
\providecommand{\url}[1]{#1}
\csname url@samestyle\endcsname
\providecommand{\newblock}{\relax}
\providecommand{\bibinfo}[2]{#2}
\providecommand{\BIBentrySTDinterwordspacing}{\spaceskip=0pt\relax}
\providecommand{\BIBentryALTinterwordstretchfactor}{4}
\providecommand{\BIBentryALTinterwordspacing}{\spaceskip=\fontdimen2\font plus
\BIBentryALTinterwordstretchfactor\fontdimen3\font minus \fontdimen4\font\relax}
\providecommand{\BIBforeignlanguage}[2]{{%
\expandafter\ifx\csname l@#1\endcsname\relax
\typeout{** WARNING: IEEEtran.bst: No hyphenation pattern has been}%
\typeout{** loaded for the language `#1'. Using the pattern for}%
\typeout{** the default language instead.}%
\else
\language=\csname l@#1\endcsname
\fi
#2}}
\providecommand{\BIBdecl}{\relax}
\BIBdecl

\bibitem{he2022collaborative}
S.~He, K.~Shi, C.~Liu, B.~Guo, J.~Chen, and Z.~Shi, ``Collaborative sensing in internet of things: A comprehensive survey,'' \emph{{IEEE} Commun. Surveys Tuts.}, 2022.

\bibitem{da2014internet}
L.~Da~Xu, W.~He, and S.~Li, ``{Internet of Things in Industries: A survey},'' \emph{{IEEE} Trans. Ind. Informat.}, vol.~10, no.~4, pp. 2233--2243, 2014.

\bibitem{sisinni2018industrial}
E.~Sisinni, A.~Saifullah, S.~Han, U.~Jennehag, and M.~Gidlund, ``{Industrial Internet of Things: Challenges, Opportunities, and Directions},'' \emph{{IEEE} Trans. Ind. Informat.}, vol.~14, no.~11, pp. 4724--4734, 2018.

\bibitem{yick2008wireless}
J.~Yick, B.~Mukherjee, and D.~Ghosal, ``Wireless sensor network survey,'' \emph{Computer networks}, vol.~52, no.~12, pp. 2292--2330, 2008.

\bibitem{duquennoy2017tsch}
S.~Duquennoy, A.~Elsts, B.~Al~Nahas, and G.~Oikonomo, ``{TSCH and 6TISCH for Contiki: Challenges, Design and Evaluation},'' in \emph{Proc. of the 13th DCOSS}.\hskip 1em plus 0.5em minus 0.4em\relax IEEE, 2017, pp. 11--18.

\bibitem{chen2016reinforcement}
H.~Chen, X.~Li, and F.~Zhao, ``{A Reinforcement Learning-based Sleep Scheduling Algorithm for Desired Area Coverage in Solar-Powered Wireless Sensor Networks},'' \emph{{IEEE} Sensors J.}, vol.~16, no.~8, pp. 2763--2774, 2016.

\bibitem{luong2019applications}
N.~C. Luong, D.~T. Hoang, S.~Gong, D.~Niyato, P.~Wang, Y.-C. Liang, and D.~I. Kim, ``{Applications of Deep Reinforcement Learning in Communications and Networking: A Survey},'' \emph{{IEEE} Commun. Surveys Tuts.}, vol.~21, no.~4, pp. 3133--3174, 2019.

\bibitem{yu2022learning}
H.~Yu and K.-W. Chin, ``{Learning Algorithms for Data Collection in RF-Charging IIoT Networks},'' \emph{{IEEE} Trans. Ind. Informat.}, vol.~19, no.~1, pp. 88--97, 2022.

\bibitem{barto2003recent}
A.~G. Barto and S.~Mahadevan, ``Recent advances in hierarchical reinforcement learning,'' \emph{Discrete event dynamic systems}, vol.~13, no. 1-2, pp. 41--77, 2003.

\bibitem{pateria2021hierarchical}
S.~Pateria, B.~Subagdja, A.-h. Tan, and C.~Quek, ``{Hierarchical Reinforcement Learning: A Comprehensive Survey},'' \emph{ACM Computing Surveys (CSUR)}, vol.~54, no.~5, pp. 1--35, 2021.

\bibitem{jurado2022survey}
F.~F. Jurado-Lasso, L.~Marchegiani, J.~F. Jurado, A.~M. Abu-Mahfouz, and X.~Fafoutis, ``{A Survey on Machine Learning Software-Defined Wireless Sensor Networks (ML-SDWSNS): Current Status and Major Challenges},'' \emph{{IEEE} Access}, vol.~10, pp. 23\,560--23\,592, 2022.

\bibitem{kim2020machine}
T.~Kim, L.~F. Vecchietti, K.~Choi, S.~Lee, and D.~Har, ``{Machine Learning for Advanced Wireless Sensor Networks: A Review},'' \emph{{IEEE} Sensors J.}, vol.~21, no.~11, pp. 12\,379--12\,397, 2020.

\bibitem{jin2016centralized}
Y.~Jin, P.~Kulkarni, J.~Wilcox, and M.~Sooriyabandara, ``A centralized scheduling algorithm for ieee 802.15. 4e tsch based industrial low power wireless networks,'' in \emph{Proc. of the 2016 IEEE WCNC}.\hskip 1em plus 0.5em minus 0.4em\relax IEEE, 2016, pp. 1--6.

\bibitem{jerbi2022msu}
W.~Jerbi, O.~Cheickhrouhou, A.~Guermazi, and H.~Trabelsi, ``Msu-tsch: A mobile scheduling updated algorithm for tsch in the internet of things,'' \emph{{IEEE} Trans. Ind. Informat.}, 2022.

\bibitem{tavakoli2018topology}
R.~Tavakoli, M.~Nabi, T.~Basten, and K.~Goossens, ``Topology management and tsch scheduling for low-latency convergecast in in-vehicle wsns,'' \emph{{IEEE} Trans. Ind. Informat.}, vol.~15, no.~2, pp. 1082--1093, 2018.

\bibitem{ojo2018throughput}
M.~O. Ojo, S.~Giordano, D.~Adami, and M.~Pagano, ``Throughput maximizing and fair scheduling algorithms in industrial internet of things networks,'' \emph{{IEEE} Trans. Ind. Informat.}, vol.~15, no.~6, pp. 3400--3410, 2018.

\bibitem{jurado2023elise}
F.~F. Jurado-Lasso, M.~Barzegaran, J.~F. Jurado, and X.~Fafoutis, ``{ELISE: A Reinforcement Learning Framework to Optimize the Slotframe Size of the TSCH Protocol in IoT Networks},'' \emph{{IEEE} Syst. J.}, 2024.

\bibitem{nguyen2019rl}
H.~Nguyen-Duy, T.~Ngo-Quynh, F.~Kojima, T.~Pham-Van, T.~Nguyen-Duc, and S.~Luongoudon, ``{RL-TSCH: A Reinforcement Learning Algorithm for Radio Scheduling in TSCH 802.15. 4e},'' in \emph{Proc. of the 2019 ICTC}.\hskip 1em plus 0.5em minus 0.4em\relax IEEE, 2019, pp. 227--231.

\bibitem{dakdouk2018reinforcement}
H.~Dakdouk, E.~Tarazona, R.~Alami, R.~F{\'e}raud, G.~Z. Papadopoulos, and P.~Maill{\'e}, ``{Reinforcement Learning Techniques for Optimized Channel Hopping in IEEE 802.15. 4-TSCH Networks},'' in \emph{Proc. of the 21st MSWiM}, 2018, pp. 99--107.

\bibitem{veisi2023enabling}
F.~Veisi, J.~Montavont, and F.~Theoleyre, ``Enabling centralized scheduling using software defined networking in industrial wireless sensor networks,'' \emph{{IEEE} Internet Things J.}, 2023.

\bibitem{kotsiou2019whitelisting}
V.~Kotsiou, G.~Z. Papadopoulos, P.~Chatzimisios, and F.~Theoleyre, ``Whitelisting without collisions for centralized scheduling in wireless industrial networks,'' \emph{{IEEE} Internet Things J.}, vol.~6, no.~3, pp. 5713--5721, 2019.

\bibitem{hajizadeh2023decentralized}
H.~Hajizadeh, M.~Nabi, and K.~Goossens, ``{Decentralized Configuration of TSCH-Based IoT Networks for Distinctive QoS: A Deep Reinforcement Learning Approach},'' \emph{{IEEE} Internet Things J.}, 2023.

\bibitem{park2020multi}
H.~Park, H.~Kim, S.-T. Kim, and P.~Mah, ``Multi-agent reinforcement-learning-based time-slotted channel hopping medium access control scheduling scheme,'' \emph{{IEEE} Access}, vol.~8, pp. 139\,727--139\,736, 2020.

\bibitem{duquennoy2015orchestra}
S.~Duquennoy, B.~Al~Nahas, O.~Landsiedel, and T.~Watteyne, ``{Orchestra: Robust Mesh Networks Through Autonomously Scheduled TSCH},'' in \emph{Proc. of the 13th Sensys}, 2015, pp. 337--350.

\bibitem{ha2022traffic}
Y.~Ha and S.-H. Chung, ``Traffic-aware 6tisch routing method for iiot wireless networks,'' \emph{{IEEE} Internet Things J.}, vol.~9, no.~22, pp. 22\,709--22\,722, 2022.

\bibitem{farag2020rea}
H.~Farag, S.~Grimaldi, M.~Gidlund, and P.~{\"O}sterberg, ``Rea-6tisch: Reliable emergency-aware communication scheme for 6tisch networks,'' \emph{{IEEE} Internet Things J.}, vol.~8, no.~3, pp. 1871--1882, 2020.

\bibitem{bommisetty2022resource}
L.~Bommisetty and T.~Venkatesh, ``{Resource Allocation in Time Slotted Channel Hopping (TSCH) Networks Based on Phasic Policy Gradient Reinforcement Learning},'' \emph{{IEEE} Internet Things J.}, vol.~19, p. 100522, 2022.

\bibitem{urke2021survey}
A.~R. Urke, {\O}.~Kure, and K.~{\O}vsthus, ``A survey of 802.15. 4 tsch schedulers for a standardized industrial internet of things,'' \emph{Sensors}, vol.~22, no.~1, p.~15, 2021.

\bibitem{shiny2022control}
S.~S.~G. Shiny, S.~S. Priya, and K.~Murugan, ``{Control Message Quenching-Based Communication Protocol for Energy Management in SDWSN},'' \emph{{IEEE} Trans. Netw. Service Manag.}, vol.~19, no.~3, pp. 3188--3201, 2022.

\bibitem{dunkels2011powertrace}
A.~Dunkels, J.~Eriksson, N.~Finne, and N.~Tsiftes, ``{Powertrace: Network-level Power Profiling for Low-power Wireless Networks},'' Swedish Institute of Computer Science, Tech. Rep. T2011:05, March 2011.

\bibitem{vilajosana2013realistic}
X.~Vilajosana, Q.~Wang, F.~Chraim, T.~Watteyne, T.~Chang, and K.~S. Pister, ``{A Realistic Energy Consumption Model for TSCH Networks},'' \emph{{IEEE} Sensors J.}, vol.~14, no.~2, pp. 482--489, 2013.

\bibitem{scanzio2020wireless}
S.~Scanzio, M.~G. Vakili, G.~Cena, C.~G. Demartini, B.~Montrucchio, A.~Valenzano, and C.~Zunino, ``{Wireless Sensor Networks and TSCH: A Compromise Between Reliability, Power Consumption, and Latency},'' \emph{{IEEE} Access}, vol.~8, pp. 167\,042--167\,058, 2020.

\bibitem{osterlind2006cross}
F.~Osterlind, A.~Dunkels, J.~Eriksson, N.~Finne, and T.~Voigt, ``{Cross-Level Sensor Network Simulation with Cooja},'' in \emph{Proc. of the 31st IEEE LCN}.\hskip 1em plus 0.5em minus 0.4em\relax IEEE, 2006, pp. 641--648.

\bibitem{oikonomou2022contiki}
G.~Oikonomou, S.~Duquennoy, A.~Elsts, J.~Eriksson, Y.~Tanaka, and N.~Tsiftes, ``The {Contiki-NG} open source operating system for next generation {IoT} devices,'' \emph{SoftwareX}, vol.~18, p. 101089, 2022.

\bibitem{chang20196tisch}
T.~Chang, M.~Vucinic, X.~Vilajosana, S.~Duquennoy, and D.~Dujovne, ``6tisch minimal scheduling function (msf),'' IETF, RFC 9033, May 2021.

\end{thebibliography}

\balance

\begin{IEEEbiography}[{\includegraphics[width=1in,height=1.25in,clip,keepaspectratio]{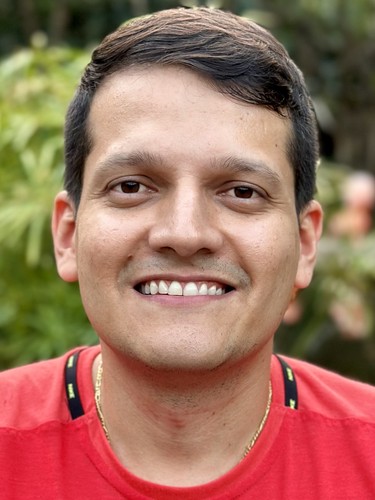}}]{F. Fernando Jurado-Lasso}
    (GS’18-M’21) received the Ph.D. degree in Engineering and the M.Eng. degree in Telecommunications Engineering both
    from The University of Melbourne, Melbourne, VIC, Australia, in 2020 and
    2015, respectively; a B.Eng. degree in Electronics Engineering in 2012 from
    the Universidad del Valle, Cali, Colombia. He is currently a postdoctoral
    researcher at the Embedded Systems Engineering (ESE) section of the
    Department of Applied Mathematics and Computer Science of the Technical
    University of Denmark (DTU Compute).
    His research interests include networked embedded systems, software-defined wireless sensor networks, machine learning, protocols and applications
    for the Internet of Things.
\end{IEEEbiography}
\begin{IEEEbiography}[{\includegraphics[width=1in,height=1.25in,clip,keepaspectratio]{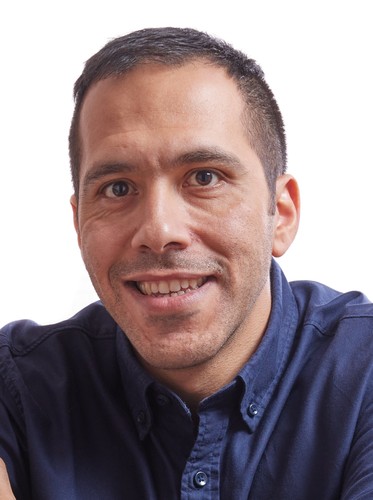}}]{Charalampos Orfanidis} received a PhD in Technology and Health from KTH Royal Institute of Technology, Stockholm, Sweden in 2020. Currently he is employed as Postdoctoral researcher at the Technical University of Denmark (DTU). His research interests span around Low-Power Wide Area Networks, Robustness, IoT and Wearables for Sports and Health.
\end{IEEEbiography}
\begin{IEEEbiography}[{\includegraphics[width=1in,height=1.25in,clip,keepaspectratio]{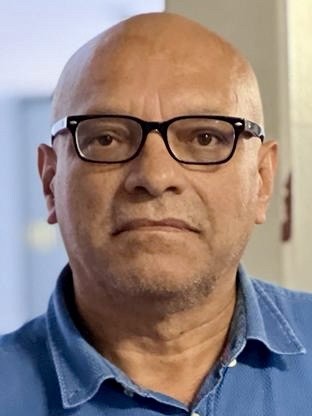}}]{J. F. Jurado}
    received the Doctorate and MSc degree in Physics both from Universidad del Valle, Cali, Colombia, in 2000 and 1986, respectively; he also holds a BSc degree in Physics from the Universidad de Nariño, Pasto, Colombia in 1984.
    He is currently a Professor with the Faculty of Engineering and Administration of the Department of Basic Science of The Universidad Nacional de Colombia Sede Palmira, Colombia. His research interests include nanomaterials, magnetic and ionic materials, nanoelectronics,  embedded systems and the Internet of Things. He is a senior member of Minciencias in Colombia.
\end{IEEEbiography}
\begin{IEEEbiography}[{\includegraphics[width=1in,height=1.25in,clip,keepaspectratio]{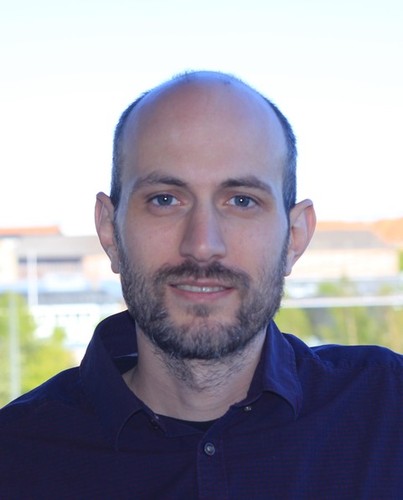}}]{Xenofon Fafoutis}
    (S'09-M'14-SM'20) received a PhD degree in Embedded Systems Engineering from the Technical University of Denmark in 2014; an MSc degree in Computer Science from the University of Crete (Greece) in 2010; and a BSc in Informatics and Telecommunications from the University of Athens (Greece) in 2007. He is currently an Associate Professor with the Embedded Systems Engineering (ESE) section of the Department of Applied Mathematics and Computer Science of the Technical University of Denmark (DTU Compute). His research interests primarily lie in Wireless Embedded Systems as an enabling technology for Digital Health, Smart Cities, and the (Industrial) Internet of Things (IoT).
\end{IEEEbiography}

\end{document}